\documentclass[onecolumn,superscriptaddress,nofootinbib,longbibliography,preprintnumbers]{revtex4-2}
\usepackage[colorlinks,linkcolor=blue,citecolor=blue,urlcolor=blue,breaklinks=true]{hyperref}
\usepackage{overpic}
\usepackage{graphicx}
\usepackage{amssymb}
\usepackage{amsmath}
\usepackage{color}
\usepackage{xcolor}
\usepackage[normalem]{ulem}
\usepackage{setspace}

\def\gTr{\textrm{gTr}}

\begin{document}

\title{Grassmann corner transfer-matrix renormalization group approach to one-dimensional fermionic models}

\author{Jian-Gang Kong}
\address{School of Physics, Renmin University of China, Beijing 100872, China}

\author{Zhi-Yuan Xie}
\email{qingtaoxie@ruc.edu.cn}
\address{School of Physics, Renmin University of China, Beijing 100872, China}
\address{Key Laboratory of Quantum State Construction and Manipulation of MoE, Renmin University of China, Beijing 100872, China}

\begin{abstract}
The strongly correlated fermions play a vital role in modern physics. For a given fermionic Hamiltonian system, the most widely used approach to explore the underlying physics is to study the wave function that incorporates Fermi-Dirac statistics, which can be obtained variationally by energy minimization or by imaginary-time evolution. In this work, we develop an accurate tensor network method for one-dimensional interacting fermionic models based on the coherent-state path-integral representation of the fermionic partition function. Employing the coherent-state representation, the partition function is effectively represented as a (1+1)-dimensional anisotropic Grassmann-valued tensor network, and the Grassmann version of the corner transfer-matrix renormalization group algorithm is developed to contract the tensor network and evaluate physical quantities. We validate our method in the one-dimensional fermionic Hubbard model with a magnetic field, where the essential features of the phase diagram in the $(\mu, B)$ plane are quantitatively captured. Our work offers a promising approach to interacting fermionic models within the framework of tensor networks.
\end{abstract}

\maketitle

\section{Introduction}

Fermionic many-body systems lie at the heart of many fundamental and technologically relevant phenomena in condensed matter physics, such as superconductivity \cite{Bednorz1986}, fractional quantum Hall effects \cite{Tsui1982}, Mott insulator \cite{MIT}, quantum spin liquid \cite{AndersonQSL}, and various topological phases of matter \cite{TopoBook}. However, accurately simulating fermionic models poses significant challenges in many-body physics, including exponential scaling with system size for exact diagonalization \cite{ExpWall}, the notorious sign problem in quantum Monte Carlo \cite{Troyer2005}, and limited correlations in approximating strongly correlated systems through weak-coupling approaches \cite{DFT}. 

Among these numerical methods, although it has its own drawbacks, the tensor network state method \cite{PEPS2004, Verstraete2008, OrusAOP,CiracRMP, Xiang2023, OrusEPJB, OrusNRP} has attracted increasing attention in recent years due to its advantages in characterizing low-entangled states and handling sign-problematic systems. In particular, by incorporating fermionic exchange statistics into the wave-function representation, the tensor network state has become an important tool for the study of strongly correlated electronic systems. For example, the swap-gate formulation \cite{Corboz2009, Corboz2010, Corboz2010a}, probably the most popular fermionic tensor network proposal \cite{Barthel2009, Kraus2009, Shi2009, Pi2010, Zhang2024, Mortier2025}, has been widely utilized for numerical simulations and provided convincing evidences of possible striped states \cite{Corboz2011, Corboz2014, Zheng2017, Ponsioen2019, Ponsioen2023} as well as superconducting states \cite{Corboz2014, Ponsioen2019, Ponsioen2023} in the two-dimensional (2D) $t$-$J$ \cite{Zhang1988} and Hubbard models \cite{Hubbard1963, Qin2022}. Nevertheless, based on the graphical projection of the 2D wave function, hosting a three-dimensional structure essentially, to a 2D plane, this swap-gate formulation is not so convenient in some situations, e.g., complex lattice structure where the graphical projection might be too complicated, and finite temperature calculation, which requires the evaluation of the partition function.

Using the coherent-state representation, Grassmann tensor networks can describe both the fermionic wavefunction and the path-integral representation of the partition function. In condensed matter physics, Gu \textit{et al.} proposed the Grassmann tensor network states \cite{Gu2010, Gu2013,Lou2015} for simulating the ground state of fermionic Hamiltonians \cite{Gu2013a, Gu2020, Miao2025,Liu2025}. Later in the particle physics community, the Grassmann tensor renormalization group was developed \cite{Levin2007, Yuya2014, Takeda2015, Sakai2017, Yoshimura2018, Akiyama2021a, Akiyama2024} to compute the partition function of relativistic lattice fermionic models. Despite their capability, Grassmann tensor networks are still relatively underexploited compared to their bosonic counterparts, and their potential has not been fully explored. In particular, the corner transfer-matrix renormalization group (CTMRG) method \cite{Nishino1996, Nishino1996a, Corboz2014, Or2009, Fishman2018, VCTMRG2022} has demonstrated supreme advantages over many coarse-graining tensor renormalization group methods in contracting 2D bosonic tensor networks, but it has not been utilized so far to showcase its power for Grassmann tensor networks. In this work, we introduce the Grassmannization of the CTMRG method and validate this approach via the one-dimensional fermionic Hubbard model \cite{Essler2005} with magnetic field in the path integral formalism \cite{Akiyama2021, Akiyama2022}. 

The rest of the paper is structured as follows. In Sec.~\ref{Sec:GTNS}, we introduce the one-dimensional Hubbard model and the Grassmann tensor network representation of its partition function in the path integral formalism. In Sec.~\ref{Sec:GCTMRG}, we firstly present a simple example of Grassmann tensor contraction as a warm-up, then the Grassmannization of the CTMRG method will be described in detail. In Sec.~\ref{Sec:Res}, we validate our Grassmann tensor network approach in the one-dimensional Hubbard model under a magnetic field. In Sec.~\ref{Sec:Summary}, we summarize and provide a prospect on the Grassmann tensor networks.

\section{Tensor network representation of fermionic partition functions} \label{Sec:GTNS}

Let us take the one-dimensional fermionic Hubbard model under a magnetic field as a concrete example. The Hamiltonian is given by
\begin{equation}\label{hubbard_H}
H = -t\sum_{\langle ij \rangle}\sum_{s=\uparrow,\downarrow} 
\left( 
\hat{c}_{is}^{\dagger}\hat{c}_{js} + h.c. 
\right) 
+ U \sum_{i}\left(
\hat{n}_{i\uparrow} - 1/2
\right)
\left(
\hat{n}_{i\downarrow} - 1/2
\right)
- \mu \sum_{i} \left(
\hat{n}_{i\uparrow} + \hat{n}_{i\downarrow} 
\right)
- B 
\sum_{i}
\left(
\hat{n}_{i\uparrow} - \hat{n}_{i\downarrow} 
\right)
\end{equation}

The $\hat{c}_{is}^{\dagger}$ and $\hat{c}_{is}$ with $s=\{\uparrow,\downarrow\}$ are the fermionic creation and annihilation operators defined at the $i$-th lattice site, and $\hat{n}_{is} \equiv \hat{c}^{\dagger}_{is}\hat{c}_{is}$ is the occupation number operator. The first term describes the electrons hopping with energy $t$ on the one-dimensional lattice, the second term is the on-site Coulomb interactions with strength $U$, the third term includes the chemical potential $\mu$ to control the electron density, and the last term describes the electrons coupled to an external magnetic field $B$. Following the standard fermionic path integral formalism \cite{ColemanBook2015}, the corresponding fermionic action for Eq.~(\ref{hubbard_H}) can be expressed as:
\begin{eqnarray} \label{action}
& & 
S\left[\psi, \bar{\psi}\right] 
= 
\sum_{n\in \Lambda_{1+1}}
\bar{\psi}(n)
\left( 
\psi(n+\hat\tau) - \psi(n) 
\right) 
- \epsilon t
\left(
\bar{\psi}(n) \psi(n+\hat\sigma) 
+ 
\bar{\psi}(n+\hat{\sigma})
\psi(n)\right)\\ \nonumber
&+&
\epsilon U 
\bar{\psi}_{\uparrow}(n)
\psi_{\uparrow}(n)
\bar{\psi}_{\downarrow}(n)
\psi_{\downarrow}(n)
- 
\epsilon\left( \mu + B + U/2 \right)
\bar{\psi}_{\uparrow}(n)
\psi_{\uparrow}(n)
-
\epsilon\left( \mu - B + U/2 \right)
\bar{\psi}_{\downarrow}(n)
\psi_{\downarrow}(n)
+
U/4
\end{eqnarray}

\begin{figure}[htbp] 
	\centering
	\includegraphics[height=7.0cm,width=14.0cm]{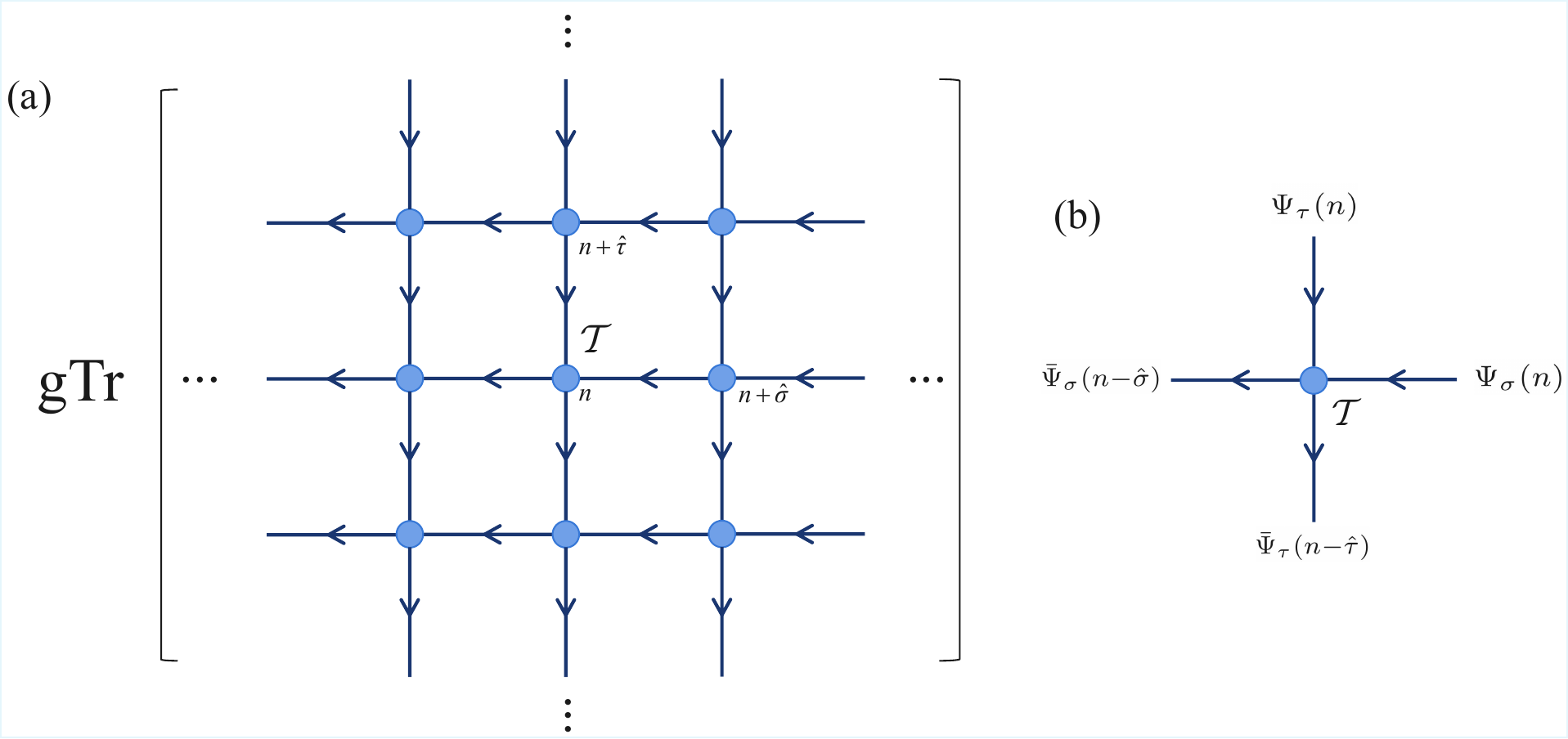}
	\caption{(a) The partition function of the one-dimensional Hubbard model represented as a Grassmann tensor network, as expressed in Eq.~(\ref{hubbard_partition_function}). (b) The local Grassmann tensor residing at the $n$-th lattice site in (a), as expressed in Eq.~(\ref{gtensor_bulk}).}
	\label{PFtns}
\end{figure}

In Eq.~(\ref{action}), $\hat{\sigma}$ ($\hat{\tau}$) denotes the unit vector in the spatial (temporal) direction of the (1+1)-dimensional lattice $\Lambda_{1+1}$, $n\equiv(n_\sigma, n_\tau)$ is the composite coordinate, and $\epsilon$ stands for the discretization (Trotter) step in the temporal (imaginary-time) direction. The $\psi(n) = \left(\psi_{\uparrow}(n),\psi_{\downarrow}(n)\right)^{T}$ and $\bar{\psi}(n)=\left(
\bar{\psi}_{\uparrow}(n),
\bar{\psi}_{\downarrow}(n)\right)$ are the two-component Grassmann-valued fields defined on the $n$-th site. $\bar{\psi}_{\uparrow,\downarrow}$ is the dual of Grassmann variable $\psi_{\uparrow,\downarrow}$, and they should be treated as strictly independent \cite{AtlandQFT}. The partition function corresponding to this Hamiltonian can be approximately represented as a 2D Grassmann tensor network \cite{Akiyama2021}:
\begin{eqnarray}\label{hubbard_partition_function}
Z 
&\approx&
\int 
\prod_{n\in \Lambda_{1+1}}
d\bar{\psi}(n)d\psi(n)
\mathrm{e}^{-S\left[\psi, \bar{\psi}\right] }
=
\gTr
\left(
\prod_{n\in \Lambda_{1+1}}
\mathcal{T}_{
\Psi_{\sigma}(n) 
\Psi_{\tau}(n)
\bar{\Psi}_{\tau}(n-\hat{\tau})
\bar{\Psi}_{\sigma}(n-\hat{\sigma})
}
\right)
\label{Eq:Zeq}
\end{eqnarray}
where the integral is over all the Grassmann variables. Similarly, the $\gTr$ stands for the Grassmann integral over all the Grassmann variables appearing in the Grassmann tensor network. An illustration is shown in Fig.~\ref{PFtns}(a). The local Grassmann tensor $\mathcal{T}$ is defined as: 
\begin{equation} \label{gtensor_bulk}
\mathcal{T}_{
\Psi_{\sigma}(n) 
\Psi_{\tau}(n)
\bar{\Psi}_{\tau}(n-\hat{\tau})
\bar{\Psi}_{\sigma}(n-\hat{\sigma})
}
=
\sum_{I_\sigma I_\sigma'=1}^{16} 
\sum_{I_\tau I_\tau'=1}^{4}
T_{I_\sigma I_\tau I_\sigma' I_\tau'}
\Psi_{\sigma}^{p(I_\sigma)}(n)
\Psi_{\tau}^{p(I_\tau)}(n)
\bar{\Psi}_{\tau}^{p(I_\tau')}(n-\hat{\tau})
\bar{\Psi}_{\sigma}^{p(I_\sigma')}(n-\hat{\sigma})
\end{equation}
Here, $\Psi_{\sigma}(n)$ and $\Psi_{\tau}(n)$ are auxiliary Grassmann variables residing on the bonds ($n$, $n + \hat{\sigma}$) and ($n$, $n + \hat{\tau}$), respectively. Conversely, $\bar{\Psi}_{\sigma}(n-\hat{\sigma})$ and $\bar{\Psi}_{\tau}(n-\hat{\tau})$ are Grassmann variables that introduced on bonds ($n - \hat{\sigma}$, $n$) and ($n - \hat{\tau}$, $n$), see Fig.~\ref{PFtns}(b). Mathematically, the Grassmann tensor $\mathcal{T}$ in Eq.~(\ref{gtensor_bulk}) represents a multilinear combination of Grassmann variables, and $T$ with commuting elements is called its coefficient tensor. The parity function $p$ yields 0 (1) if the index belongs to the Grassmann-even (odd) sector. For a more detailed introduction to Grassmann tensor operations, such as the Grassmann contraction, fusion,  decomposition, etc., interested readers can refer to Refs.~\cite{Akiyama2024}. For action expressed in Eq.~(\ref{action}), the coefficient tensor $T$ appearing in Eq.~(\ref{gtensor_bulk}) is determined as:
\begin{eqnarray} \label{coef_tensor_bulk}
& & 
T_{I_\sigma I_\tau I_\sigma' I_\tau'}
\xleftarrow{fuse}
T_{
	(i_{\sigma\uparrow} i_{\sigma\downarrow}
	j_{\sigma\uparrow} j_{\sigma\downarrow})
	(i_{\tau\uparrow} i_{\tau\downarrow})
	(i_{\sigma\uparrow}^{'} i_{\sigma\downarrow}^{'}
	j_{\sigma\uparrow}^{'} j_{\sigma\downarrow}^{'})
	(i_{\tau\uparrow}' i_{\tau\downarrow}')
}
\\ \nonumber
&=&
\sum_{i_{\sigma\uparrow} i_{\sigma\downarrow}} 
\sum_{i_{\sigma\uparrow}^{'} i_{\sigma\downarrow}^{'}} 
\sum_{j_{\sigma\uparrow} j_{\sigma\downarrow}}
\sum_{j_{\sigma\uparrow}^{'} j_{\sigma\downarrow}^{'}}
\sum_{i_{\tau\uparrow} i_{\tau\downarrow}} 
\sum_{i_{\tau\uparrow}^{'} i_{\tau\downarrow}^{'}} 
e^{-U\epsilon/4}
\times
(-1)^{i_{\sigma\uparrow}+i_{\sigma\downarrow}}
\times 
(\sqrt{\epsilon t})^{\sum_{s=\uparrow\downarrow}(i_{\sigma s} + i_{\sigma s}^{'} + j_{\sigma s} + j_{\sigma s}^{'})}
\times
(-1)^R
\\ \nonumber
&\times&
(
\delta_{
	i_{\tau\downarrow}^{'} +
	i_{\sigma\downarrow}^{'} +  
	j_{\sigma\downarrow}, 1} 
\delta_{
	i_{\tau\downarrow}
	i_{\sigma\downarrow}
	j_{\sigma\downarrow}^{'}, 1} 
\delta_{
	i_{\tau\uparrow}^{'} +
	i_{\sigma\uparrow}^{'} + 
	j_{\sigma\uparrow}, 1}
\delta_{
	i_{\tau\uparrow} +
	i_{\sigma\uparrow} +
	j_{\sigma\uparrow}^{'}, 1} 
\\ \nonumber
&-& 
\left[
\epsilon\left( \mu + B + U/2 \right)
+
1
\right]
\delta_{
	i_{\tau\downarrow}^{'} +
	i_{\sigma\downarrow}^{'} +  
	j_{\sigma\downarrow}, 1} 
\delta_{
	i_{\tau\downarrow}
	i_{\sigma\downarrow}
	j_{\sigma\downarrow}^{'}, 1} 
\delta_{
	i_{\tau\uparrow}^{'} +
	i_{\sigma\uparrow}^{'} + 
	j_{\sigma\uparrow}, 0}
\delta_{
	i_{\tau\uparrow} +
	i_{\sigma\uparrow} +
	j_{\sigma\uparrow}^{'}, 0} 
\\ \nonumber
&-& 
\left[
\epsilon\left( \mu - B + U/2 \right)
+
1
\right]
\delta_{
	i_{\tau\downarrow}^{'} +
	i_{\sigma\downarrow}^{'} +  
	j_{\sigma\downarrow}, 0} 
\delta_{
	i_{\tau\downarrow}
	i_{\sigma\downarrow}
	j_{\sigma\downarrow}^{'}, 0} 
\delta_{
	i_{\tau\uparrow}^{'} +
	i_{\sigma\uparrow}^{'} + 
	j_{\sigma\uparrow}, 1}
\delta_{
	i_{\tau\uparrow} +
	i_{\sigma\uparrow} +
	j_{\sigma\uparrow}^{'}, 1} 
\\ \nonumber
&+& 
\left(
\left[
\epsilon
\left( \mu + B + U/2 \right)
+
1
\right]
\left[
\epsilon
\left( \mu - B + U/2 \right)
+
1
\right]
-
\epsilon U
\right)
\delta_{
	i_{\tau\downarrow}^{'} +
	i_{\sigma\downarrow}^{'} +  
	j_{\sigma\downarrow}, 0} 
\delta_{
	i_{\tau\downarrow}
	i_{\sigma\downarrow}
	j_{\sigma\downarrow}^{'}, 0} 
\delta_{
	i_{\tau\uparrow}^{'} +
	i_{\sigma\uparrow}^{'} + 
	j_{\sigma\uparrow}, 0}
\delta_{
	i_{\tau\uparrow} +
	i_{\sigma\uparrow} +
	j_{\sigma\uparrow}^{'}, 0} 
)
\end{eqnarray}
where the $\xleftarrow{fuse}$ denotes the fact that the rank-4 tensor $T$ is obtained by fusing its rank-12 counterpart with $i_{xs},i_{xs}',j_{xs},j_{xs}'=\{0,1\}$, $x = \{\sigma,\tau\}$, and $s=\{\uparrow,\downarrow\}$. $(-1)^R$ is the sign factor responsible for the reordering of Grassmann variables. A detailed derivation of Eq.~(\ref{coef_tensor_bulk}) is given in Appendix \cite{appendix}. 

Now we have represented the partition function of the one-dimensional Hubbard model, expressed in Eq.~(\ref{hubbard_H}), as the $\gTr$ of a Grassmann tensor network approximately, which is expressed in Eq.~(\ref{hubbard_partition_function}) and illustrated in Fig.~\ref{PFtns}(a). In this work, we are interested in exploring the phase diagram of the one-dimensional Hubbard model in the $(\mu, B)$ plane, which can be characterized by physical quantities such as particle number $p = \langle \hat{n}_{\uparrow} \rangle + \langle \hat{n}_{\downarrow} \rangle$, magnetization $m = (\langle \hat{n}_{\uparrow} \rangle - \langle \hat{n}_{\downarrow} \rangle)/2$, and the double occupancy $d = \langle \hat{n}_{\uparrow}\hat{n}_{\downarrow} \rangle$. In the context of the tensor networks, those quantities can be evaluated through the so-called impurity method \cite{HHZhaoPRB2016}:
\begin{eqnarray} \label{nu_impurity}
\langle \hat{n}_{\uparrow}(l) \rangle
\approx
\dfrac{
\left(
\int 
\prod_{n\in \Lambda_{1+1}}
d\bar{\psi}(n)d\psi(n)
\right)
\bar{\psi}_{\uparrow}(l)
\psi_{\uparrow}(l)
\mathrm{e}^{-S\left[\psi, \bar{\psi}\right]}
}{Z}
=
\dfrac{\textrm{gTr}
\left(
\cdots
\mathcal{T}^{n_{\uparrow}}_{
\Psi_{\sigma}
\Psi_{\tau}
\bar{\Psi}_{\tau}
\bar{\Psi}_{\sigma}
}
\cdots
\right)}{Z}
\end{eqnarray}
\begin{eqnarray} \label{nd_impurity}
\langle \hat{n}_{\downarrow}(l) \rangle
\approx
\dfrac{
\left(
\int 
\prod_{n\in \Lambda_{1+1}}
d\bar{\psi}(n)
d\psi(n)
\right)
\bar{\psi}_{\downarrow}(l)
\psi_{\downarrow}(l)
\mathrm{e}^{-S\left[\psi, \bar{\psi}\right]}
}{Z}
=
\dfrac{\textrm{gTr}
\left(
\cdots
\mathcal{T}^{n_{\downarrow}}_{
\Psi_{\sigma}
\Psi_{\tau}
\bar{\Psi}_{\tau}
\bar{\Psi}_{\sigma}
}
\cdots
\right)}{Z}
\end{eqnarray}
\begin{eqnarray} \label{nund_impurity}
\langle \hat{n}_{\uparrow}(l)\hat{n}_{\downarrow}(l)\rangle
\approx
\dfrac{
\left(
\int 
\prod_{n\in \Lambda_{1+1}}
d\bar{\psi}(n)d\psi(n)
\right)
\bar{\psi}_{\uparrow}(l)
\psi_{\uparrow}(l)
\bar{\psi}_{\downarrow}(l)
\psi_{\downarrow}(l)
\mathrm{e}^{-S\left[\psi, \bar{\psi}\right]}
}
{Z}
=
\dfrac{\textrm{gTr}
\left(
\cdots
\mathcal{T}^{d}_{
\Psi_{\sigma}
\Psi_{\tau}
\bar{\Psi}_{\tau}
\bar{\Psi}_{\sigma}
}
\cdots
\right)}{Z}
\end{eqnarray}
Here, for simplicity, we have omitted the site labels of the Grassmann subscripts in $\mathcal{T}$. For clarity, we emphasize that the occupation number operator $\hat{n}$ and the lattice site label $n$ refer to different quantities and are not to be confused. In the numerators of Eqs.~(\ref{nu_impurity}-\ref{nund_impurity}), the Grassmann tensor defined at the $l$-th lattice site is replaced by the impurity Grassmann tensors corresponding to $\hat{n}_{\uparrow}$, $\hat{n}_{\downarrow}$ and $\hat{n}_{\uparrow}\hat{n}_{\downarrow}$, respectively. To be specific, the corresponding coefficients of the three impurity tensors are defined in the following:
\begin{eqnarray} \label{nu_impurity_coefficient}
& &
T_{I_\sigma I_\tau I_\sigma' I_\tau'}^{n_{\uparrow}}
\xleftarrow{fuse}
T_{
	(i_{\sigma\uparrow} i_{\sigma\downarrow}
	j_{\sigma\uparrow} j_{\sigma\downarrow})
	(i_{\tau\uparrow} i_{\tau\downarrow})
	(i_{\sigma\uparrow}^{'} i_{\sigma\downarrow}^{'}
	j_{\sigma\uparrow}^{'} j_{\sigma\downarrow}^{'})
	(i_{\tau\uparrow}' i_{\tau\downarrow}')
}^{n_{\uparrow}}
\\ \nonumber
&=&
\sum_{i_{\sigma\uparrow} i_{\sigma\downarrow}} 
\sum_{i_{\sigma\uparrow}^{'} i_{\sigma\downarrow}^{'}} 
\sum_{j_{\sigma\uparrow} j_{\sigma\downarrow}}
\sum_{j_{\sigma\uparrow}^{'} j_{\sigma\downarrow}^{'}}
\sum_{i_{\tau\uparrow} i_{\tau\downarrow}} 
\sum_{i_{\tau\uparrow}^{'} i_{\tau\downarrow}^{'}} 
e^{-U\epsilon/4}
\times
(-1)^{i_{\sigma\uparrow}+i_{\sigma\downarrow}}
\times 
(\sqrt{\epsilon t})^{\sum_{s=\uparrow\downarrow}(i_{\sigma s} + i_{\sigma s}^{'} + j_{\sigma s} + j_{\sigma s}^{'})}
\times
(-1)^R
\\ \nonumber
&\times&
(
-
\delta_{
	i_{\tau\downarrow}^{'} +
	i_{\sigma\downarrow}^{'} +  
	j_{\sigma\downarrow}, 1} 
\delta_{
	i_{\tau\downarrow}
	i_{\sigma\downarrow}
	j_{\sigma\downarrow}^{'}, 1} 
\delta_{
	i_{\tau\uparrow}^{'} +
	i_{\sigma\uparrow}^{'} + 
	j_{\sigma\uparrow}, 0}
\delta_{
	i_{\tau\uparrow} +
	i_{\sigma\uparrow} +
	j_{\sigma\uparrow}^{'}, 0} 
\\ \nonumber
&+&
\left[
\epsilon\left( \mu - B + U/2 \right)
+
1
\right]
\delta_{
	i_{\tau\downarrow}^{'} +
	i_{\sigma\downarrow}^{'} +  
	j_{\sigma\downarrow}, 0} 
\delta_{
	i_{\tau\downarrow}
	i_{\sigma\downarrow}
	j_{\sigma\downarrow}^{'}, 0} 
\delta_{
	i_{\tau\uparrow}^{'} +
	i_{\sigma\uparrow}^{'} + 
	j_{\sigma\uparrow}, 0}
\delta_{
	i_{\tau\uparrow} +
	i_{\sigma\uparrow} +
	j_{\sigma\uparrow}^{'}, 0} 
)
\end{eqnarray}
\begin{eqnarray} \label{nd_impurity_coefficient}
& &
T_{I_\sigma I_\tau I_\sigma' I_\tau'}^{n_{\downarrow}}
\xleftarrow{fuse}
T_{
	(i_{\sigma\uparrow} i_{\sigma\downarrow}
	j_{\sigma\uparrow} j_{\sigma\downarrow})
	(i_{\tau\uparrow} i_{\tau\downarrow})
	(i_{\sigma\uparrow}^{'} i_{\sigma\downarrow}^{'}
	j_{\sigma\uparrow}^{'} j_{\sigma\downarrow}^{'})
	(i_{\tau\uparrow}' i_{\tau\downarrow}')
}^{n_{\downarrow}}
\\ \nonumber
&=&
\sum_{i_{\sigma\uparrow} i_{\sigma\downarrow}} 
\sum_{i_{\sigma\uparrow}^{'} i_{\sigma\downarrow}^{'}} 
\sum_{j_{\sigma\uparrow} j_{\sigma\downarrow}}
\sum_{j_{\sigma\uparrow}^{'} j_{\sigma\downarrow}^{'}}
\sum_{i_{\tau\uparrow} i_{\tau\downarrow}} 
\sum_{i_{\tau\uparrow}^{'} i_{\tau\downarrow}^{'}} 
e^{-U\epsilon/4}
\times
(-1)^{i_{\sigma\uparrow}+i_{\sigma\downarrow}}
\times 
(\sqrt{\epsilon t})^{\sum_{s=\uparrow\downarrow}(i_{\sigma s} + i_{\sigma s}^{'} + j_{\sigma s} + j_{\sigma s}^{'})}
\times
(-1)^R
\\ \nonumber
&\times&
(
-
\delta_{
	i_{\tau\downarrow}^{'} +
	i_{\sigma\downarrow}^{'} +  
	j_{\sigma\downarrow}, 0} 
\delta_{
	i_{\tau\downarrow}
	i_{\sigma\downarrow}
	j_{\sigma\downarrow}^{'}, 0} 
\delta_{
	i_{\tau\uparrow}^{'} +
	i_{\sigma\uparrow}^{'} + 
	j_{\sigma\uparrow}, 1}
\delta_{
	i_{\tau\uparrow} +
	i_{\sigma\uparrow} +
	j_{\sigma\uparrow}^{'}, 1} 
\\ \nonumber
&+&
\left[
\epsilon\left( \mu + B + U/2 \right)
+
1
\right]
\delta_{
	i_{\tau\downarrow}^{'} +
	i_{\sigma\downarrow}^{'} +  
	j_{\sigma\downarrow}, 0} 
\delta_{
	i_{\tau\downarrow}
	i_{\sigma\downarrow}
	j_{\sigma\downarrow}^{'}, 0} 
\delta_{
	i_{\tau\uparrow}^{'} +
	i_{\sigma\uparrow}^{'} + 
	j_{\sigma\uparrow}, 0}
\delta_{
	i_{\tau\uparrow} +
	i_{\sigma\uparrow} +
	j_{\sigma\uparrow}^{'}, 0})
\end{eqnarray}
\begin{eqnarray} \label{nund_impurity_coefficient}
& &
T_{I_\sigma I_\tau I_\sigma' I_\tau'}^{d}
\xleftarrow{fuse}
T_{
	(i_{\sigma\uparrow} i_{\sigma\downarrow}
	j_{\sigma\uparrow} j_{\sigma\downarrow})
	(i_{\tau\uparrow} i_{\tau\downarrow})
	(i_{\sigma\uparrow}^{'} i_{\sigma\downarrow}^{'}
	j_{\sigma\uparrow}^{'} j_{\sigma\downarrow}^{'})
	(i_{\tau\uparrow}' i_{\tau\downarrow}')
}^{d}
\\ \nonumber
&=&
\sum_{i_{\sigma\uparrow} i_{\sigma\downarrow}} 
\sum_{i_{\sigma\uparrow}^{'} i_{\sigma\downarrow}^{'}} 
\sum_{j_{\sigma\uparrow} j_{\sigma\downarrow}}
\sum_{j_{\sigma\uparrow}^{'} j_{\sigma\downarrow}^{'}}
\sum_{i_{\tau\uparrow} i_{\tau\downarrow}} 
\sum_{i_{\tau\uparrow}^{'} i_{\tau\downarrow}^{'}} 
e^{-U\epsilon/4}
\times
(-1)^{i_{\sigma\uparrow}+i_{\sigma\downarrow}}
\times 
(\sqrt{\epsilon t})^{\sum_{s=\uparrow\downarrow}(i_{\sigma s} + i_{\sigma s}^{'} + j_{\sigma s} + j_{\sigma s}^{'})}
\times
(-1)^R
\\ \nonumber
&\times&
\delta_{
	i_{\tau\downarrow}^{'} +
	i_{\sigma\downarrow}^{'} +  
	j_{\sigma\downarrow}, 0} 
\delta_{
	i_{\tau\downarrow}
	i_{\sigma\downarrow}
	j_{\sigma\downarrow}^{'}, 0} 
\delta_{
	i_{\tau\uparrow}^{'} +
	i_{\sigma\uparrow}^{'} + 
	j_{\sigma\uparrow}, 0}
\delta_{
	i_{\tau\uparrow} +
	i_{\sigma\uparrow} +
	j_{\sigma\uparrow}^{'}, 0}
\end{eqnarray}

\section{Grassmann CTMRG} \label{Sec:GCTMRG}

We now turn to the Grassmannization of the CTMRG method. However, before diving into the details, let us consider the Grassmann tensor contraction, the most basic operation in all Grassmann tensor network methods, as a warm-up. To make it concrete, let us consider the contraction between two neighboring rank-4 Grassmann tensors, $\mathcal{T}^u$ and $\mathcal{T}^d$, along the imaginary-time direction, as sketched in Fig.~\ref{g_contract}(a). Specifically, the two Grassmann tensors are defined in the following:
\begin{figure}[htbp] 
	\centering
	\includegraphics[height=7.0cm,width=11.5cm]{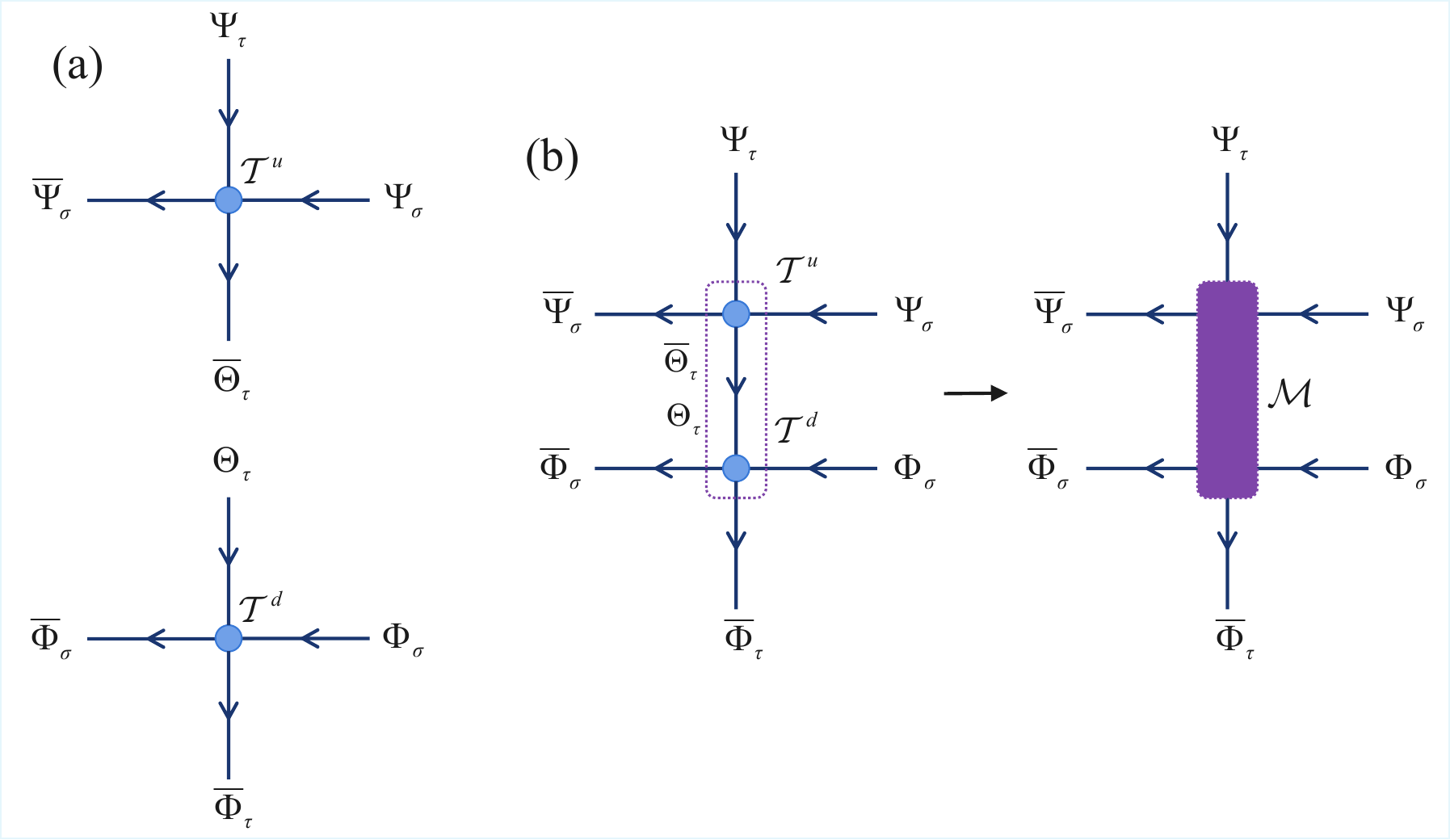}
	\caption{(a) The Grassmann tensors $\mathcal{T}^{u}$ and $\mathcal{T}^{d}$. (b) The Grassmann contraction of $\mathcal{T}^{u}$ and $\mathcal{T}^{d}$.}
	\label{g_contract}
\end{figure}
\begin{eqnarray} \label{grassmann_Tu}
\mathcal{T}^{u}_{
	\Psi_{\sigma}
    \Psi_{\tau}
    \bar{\Psi}_{\sigma}
    \bar{\Theta}_{\tau}
}
=
\sum_{I_{\sigma}I_{\tau}I_{\sigma}'I_{\tau}'}
T^{u}_{I_{\sigma}I_{\tau}I_{\sigma}'I_{\tau}'}
\Psi_{\sigma}^{p(I_{\sigma})}
\Psi_{\tau}^{p(I_{\tau})}
\bar{\Psi}_{\sigma}^{p(I_{\sigma}')}
\bar{\Theta}_{\tau}^{p(I_{\tau}')}
\end{eqnarray}
\begin{eqnarray} \label{grassmann_Td}
\mathcal{T}^{d}_{
	\Phi_{\sigma}
	\Theta_{\tau}
	\bar{\Phi}_{\sigma}
	\bar{\Phi}_{\tau}
}
=
\sum_{J_{\sigma}J_{\tau}J_{\sigma}'J_{\tau}'}
T^{d}_{J_{\sigma}J_{\tau}J_{\sigma}'J_{\tau}'}
\Phi_{\sigma}^{p(J_{\sigma})}
\Theta_{\tau}^{p(J_{\tau})}
\bar{\Phi}_{\sigma}^{p(J_{\sigma}')}
\bar{\Phi}_{\tau}^{p(J_{\tau}')}
\end{eqnarray}
As expressed in Eqs.~(\ref{grassmann_Tu}-\ref{grassmann_Td}) and shown in Fig.~\ref{g_contract}(a), an outgoing (incoming) arrow on a link indicates the related tensor index is associated with a dual (non-dual) Grassmann variable. It is important to distinguish the types of Grassmann variables, since a Grassmann variable 
$\Theta$ should always be placed to the left of its dual variable $\bar\Theta$ when performing Grassmann contraction \cite{Akiyama2021a} according to:
\begin{eqnarray} \label{gcontract}
\int_{\bar{\Theta}_{\tau}\Theta_{\tau}}
\Theta_{\tau}^{p(J_{\tau})}
\bar{\Theta}_{\tau}^{p(I_{\tau}')}
=
\delta_{J_{\tau},I_{\tau}'}
\delta_{p(J_{\tau}),p(I_{\tau}')}
,
\qquad\textrm{with}\quad
\int_{\bar{\Theta}_{\tau}\Theta_{\tau}}
\equiv
\left(
\int d\bar{\Theta}_{\tau}d\Theta_{\tau}\mathrm{e}^{-\bar{\Theta}_{\tau}\Theta_{\tau}}
\right)
\end{eqnarray}
where $\delta_{J_{\tau},I_{\tau}'} \delta_{p(J_{\tau}),p(I_{\tau}')}$ would lead to the contraction of corresponding coefficient tensors. Now we can obtain a rank-6 Grassmann tensor $\mathcal{M}$ by performing the Grassmann contraction of $\mathcal{T}^{u}$ and $\mathcal{T}^{d}$ as illustrated in Fig.~\ref{g_contract}(b):
\begin{eqnarray} \label{grassmann_M}
\mathcal{M}_{
	\Psi_{\sigma}
	\Phi_{\sigma}
	\Psi_{\tau}
	\bar{\Phi}_{\sigma}
	\bar{\Psi}_{\sigma}
	\bar{\Phi}_{\tau}
}
&=&
\int_{\bar{\Theta}_{\tau}\Theta_{\tau}}
\mathcal{T}^{u}_{
	\Psi_{\sigma}
	\Psi_{\tau}
	\bar{\Psi}_{\sigma}
	\bar{\Theta}_{\tau}
}
\mathcal{T}^{d}_{
	\Phi_{\sigma}
	\Theta_{\tau}
	\bar{\Phi}_{\sigma}
	\bar{\Phi}_{\tau}
}
\\ 
&=& \nonumber
\sum_{I_{\sigma}J_{\sigma}I_{\tau}I_{\sigma}'J_{\sigma}'J_{\tau}'}
M_{
	I_{\sigma}J_{\sigma}
	I_{\tau}
	I_{\sigma}'J_{\sigma}'
	J_{\tau}'}
\Psi_{\sigma}^{p(I_{\sigma})}
\Phi_{\sigma}^{p(J_{\sigma})}
\Psi_{\tau}^{p(I_{\tau})}
\bar{\Phi}_{\sigma}^{p(J_{\sigma}')}
\bar{\Psi}_{\sigma}^{p(I_{\sigma}')}
\bar{\Phi}_{\tau}^{p(J_{\tau}')}
\end{eqnarray}
The coefficient tensor $M$ is obtained from the ordinary tensor contraction of coefficient tensors $T^{u}$ and $T^{d}$, but multiplied with a non-trivial fermionic sign factor, i.e., %
\begin{eqnarray} \label{M_tensor}
M_{
	(I_{\sigma}J_{\sigma})
	I_{\tau}
	(I_{\sigma}'J_{\sigma}')
	J_{\tau}'}
=
\sum_{K_{\tau}}
T^{u}_{I_{\sigma}I_{\tau}I_{\sigma}'K_{\tau}}
T^{d}_{J_{\sigma}K_{\tau}J_{\sigma}'J_{\tau}'}
\times
(-1)^{S\left(
	I_{\sigma},
	I_{\tau},
	I_{\sigma}',
	J_{\sigma},
	J_{\sigma}',
	J_{\tau}',
	K_{\tau}
	\right)}
\end{eqnarray}

The sign factor $(-1)^{S}$ in Eq.~(\ref{M_tensor}) is responsible for two things: moving the Grassmann variable $\bar{\Theta}_{\tau}$ to the right-hand side of $\Theta_{\tau}$ for performing the Grassmann integral in Eq.~(\ref{gcontract}), and rearranging the rest Grassmann variables according to the order specified by the subscript of the new Grassmann tensor $\mathcal{M}$. The explicit form of the sign factor is derived in Appendix \cite{appendix}. In fact, it is unnecessary to derive the sign factors for each Grassmann contraction case by case; in practice, one can design a simple computer program to handle them in a unified way, as done in Ref.~\cite{Yosprakob2023}. As in bosonic tensor networks, when summing two given Grassmann tensors over several indices, one can always perform the contraction in two steps: (1) arrange the indices requiring summation together and put them in their appropriate places, i.e., the end of the multiplicand's indices and the beginning of the multiplier's indices; (2) do the summation over the indices and perform the corresponding Grassmann integral according to Eq.~(\ref{gcontract}). Since the parity of each index is known, the exchange of any two neighboring indices, e.g., $i$ and $j$, would contribute a sign $(-1)^{p(i)\times p(j)}$, and the total sign stemming from step (1) can be obtained by multiplying all these factors involved one by one. This can be achieved effectively by designing a nested loop that counts the exchanges involved.

After gaining some familiarity with the Grassmann notation, let us introduce the Grassmann CTMRG method. The CTMRG \cite{Nishino1996,Nishino1996a} is one of the most accurate methods to contract a two-dimensional tensor network in the thermodynamic limit. It was originally tailored for models in statistical physics and later generalized to evaluate physical quantities of condensed-matter systems using the iPEPS ansatz \cite{Or2009,Corboz2014,Fishman2018}. Recently, it has been made variational and more efficient in symmetric tensor networks \cite{VCTMRG2022}, even to other lattices than square \cite{HoneyCTMRG}. In the following, we describe a Grassmann version of the so-called directional \cite{Corboz2014}, or asymmetric \cite{Fishman2018}, CTMRG, which can be applied to tensor networks with unit cells of arbitrary size and demonstrates good overall performance in most cases. 
\begin{figure}[htbp] 
	\centering
	\includegraphics[height=5.8cm,width=10.2cm]{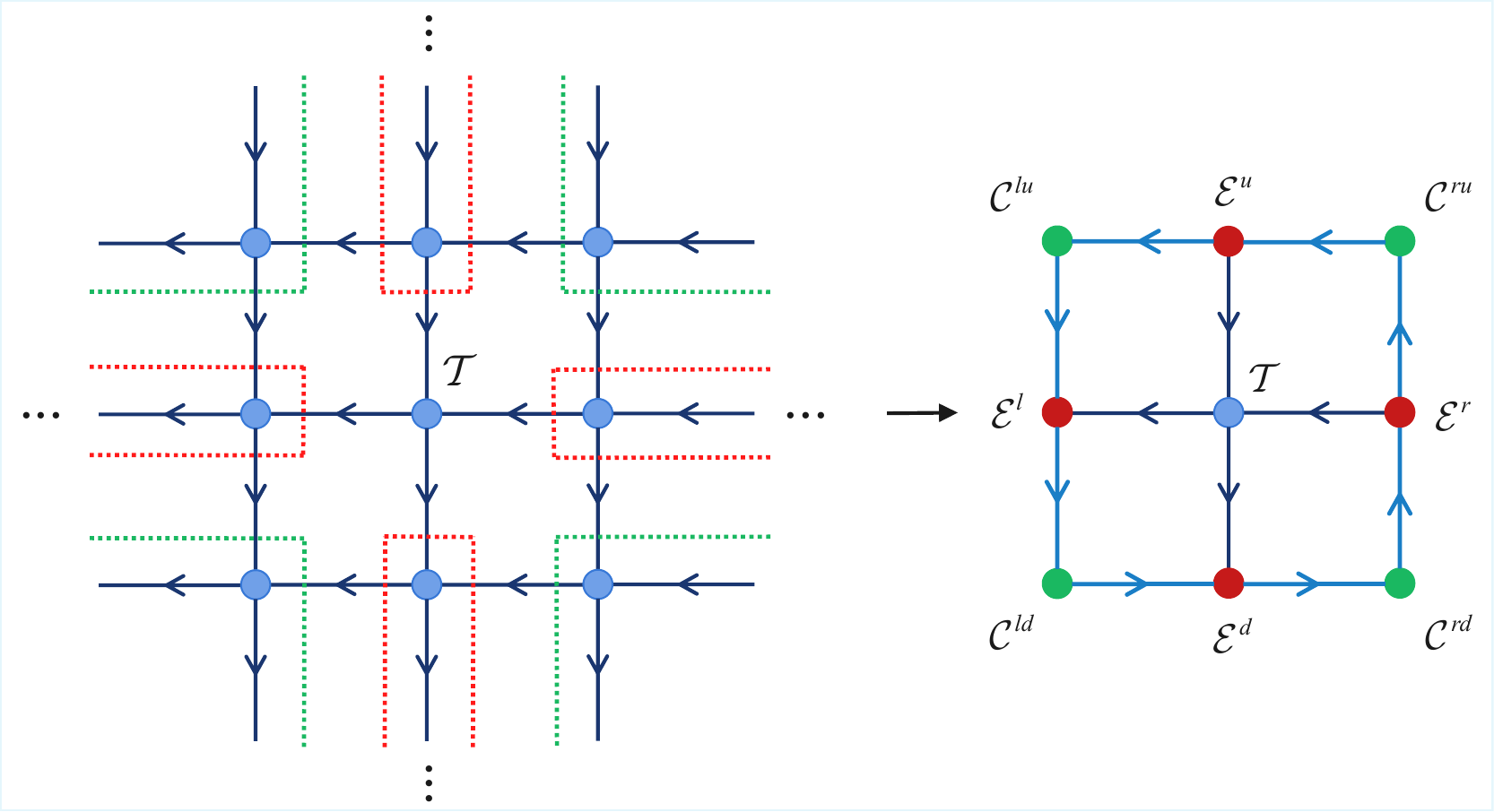}
	\caption{A Grassmann tensor network defined on a square lattice is approximated by a finite cluster of Grassmann tensors, where four corner matrices $\{\mathcal{C}\}$ and four edge tensors $\{\mathcal{E}\}$ are iteratively determined to represent the effective environments of $\mathcal{T}$.}
	\label{ctmrg_env}
\end{figure}

Suppose we are considering a 2D Grassmann tensor network generated by the uniform rank-4 Grassmann tensor $\mathcal{T}$, introduced in Eqs.~(\ref{gtensor_bulk})-(\ref{coef_tensor_bulk}). The Grassmann CTMRG aims to find four Grassmann corner matrices $\mathcal{C}^{lu},\mathcal{C}^{ru},\mathcal{C}^{ld},\mathcal{C}^{rd}$ to approximate the quadrants surrounding the bulk tensor $\mathcal{T}$, and four rank-3 Grassmann edge tensors $\mathcal{E}^{l},\mathcal{E}^{r},\mathcal{E}^{u},\mathcal{E}^{d}$ to approximate corresponding half-infinite rows and columns, as we can see in Fig.~\ref{ctmrg_env}. The accuracy is controlled by the virtual bond dimension of the environment tensors (namely the corners and edges), denoted as $\chi$ in this work. Starting from some prepared $\{\mathcal{C}\}$ and $\{\mathcal{E}\}$, which can be randomly initialized, the Grassmann CTMRG algorithm optimizes them iteratively by performing \textit{moves} in four directions one by one alternatively until convergence is reached. Here, we would outline the downward move of the procedure. 

\begin{figure}[htbp] 
	\centering
	\includegraphics[height=7.8cm,width=10.5cm]{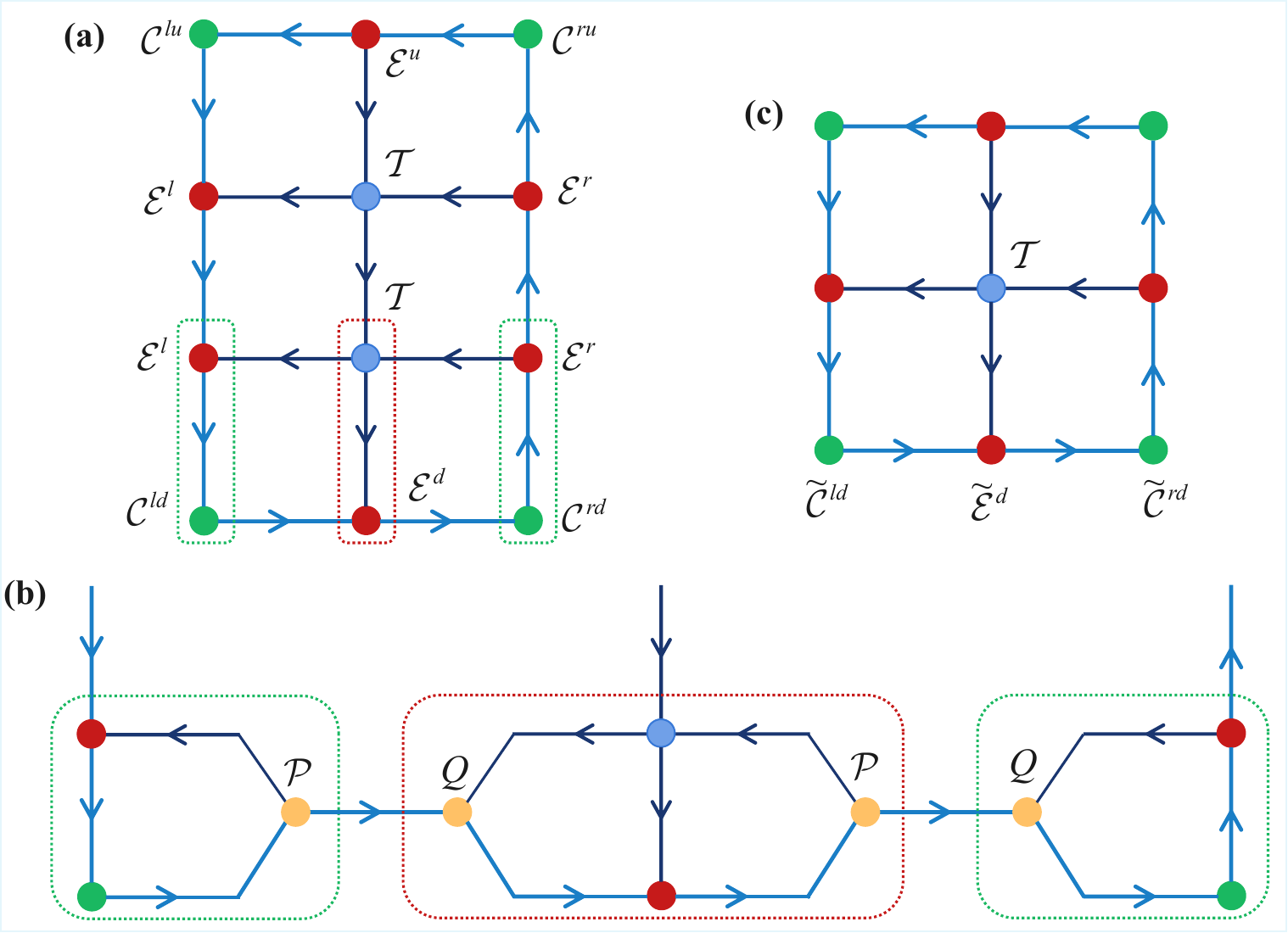}
	\caption{The downward move of the Grassmann CTMRG method. The sizes of $\mathcal{P}$ and $\mathcal{Q}$ are $\chi D\times \chi$ and $\chi \times \chi D$, respectively.}
	\label{ctmrg_renormalize}
\end{figure}

As illustrated in Fig.~\ref{ctmrg_renormalize}, the downward move would eventually update the three downward environment tensors, i.e., $\mathcal{C}^{ld}$, $\mathcal{E}^{d}$, and $\mathcal{C}^{rd}$. It starts by inserting a single row into the effective $3\times 3$ cluster in Fig.~\ref{ctmrg_env}, and subsequently absorbs the inserted tensors into the downward environments. Suppose each index of the coefficient tensor of $T$ has a common dimension $D$, then the new environments will have a dimension $\chi D$. In order to perform truncation of the enlarged dimension to the original $\chi$, a pair of Grassmann isometric tensors $\mathcal{P}$ and $\mathcal{Q}$ whose coefficient tensors are of size $\chi D\times \chi$ and $\chi \times \chi D$, respectively, is constructed and inserted at the corresponding links for contractions. And this completes a single step of the downward move. As a matter of fact, the steps described here are exactly the same as the usual bosonic CTMRG, except that we have replaced all the tensor contractions by Grassmann contractions described at the beginning of this section. However, non-trivial Grassmann sign factors would play a vital role in constructing the isometric tensors and performing contractions, which we would discuss in the following. 

\begin{figure}[htbp] 
	\centering
	\includegraphics[height=5.8cm,width=13.4cm]{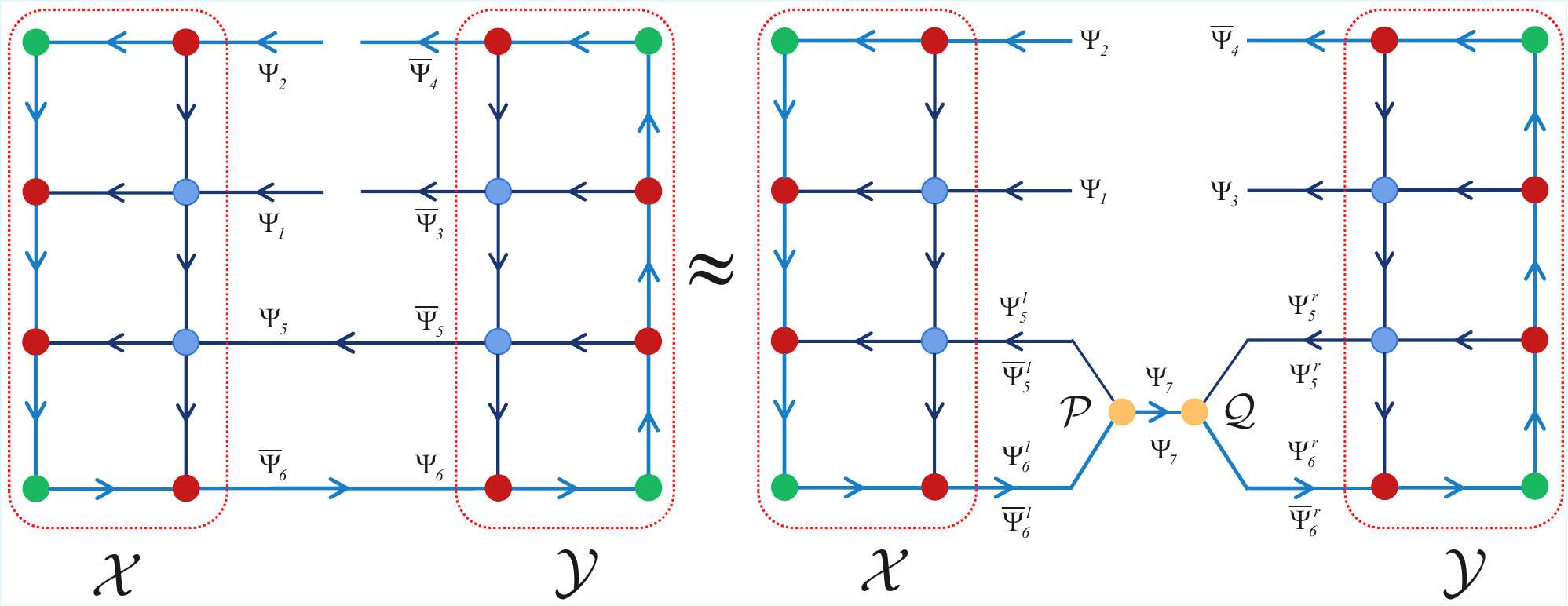}
	\caption{The Grassmann tensor equation satisfied by the Grassmann isometries $\mathcal{P}$ and $\mathcal{Q}$, i.e., their coefficients are required to construct a low-rank approximation of $W$ in Eq.~(\ref{Wdef}), or equivalently, $\mathcal{P}$ and $\mathcal{Q}$ constitute the truncated SVD of $\mathcal{W}$ as expressed in Eq.~(\ref{goal}).}
	\label{ctmrg_isometry}
\end{figure}

To construct the isometric tensors $\mathcal{P}$ and $\mathcal{Q}$, we consider a $4\times 4$ cluster with a $2\times 2~\mathcal{T}$-cluster placed in the center, as illustrated in Fig.~\ref{ctmrg_isometry}. We have also explicitly written the Grassmann variables associated with each index for clarity. Firstly, a rank-4 Grassmann tensor $\mathcal{W}$ is obtained by Grassmann contraction:
\begin{eqnarray}
\mathcal{W}_{\Psi_{1}\Psi_{2};\bar{\Psi}_{3}\bar{\Psi}_{4}} = 
\int_{\bar{\Psi}_{5}\Psi_{5}}
\int_{\bar{\Psi}_{6}\Psi_{6}}
\mathcal{X}_{\Psi_{1}\Psi_{2};\Psi_{5}\bar{\Psi}_{6}}
\mathcal{Y}_{\Psi_{6}\bar{\Psi}_{5};\bar{\Psi}_{3}\bar{\Psi}_{4}}
\label{Gcontract}
\end{eqnarray}
The coefficient tensor $W$ can be computed after performing the Grassmann contraction between $\mathcal{X}$ and $\mathcal{Y}$ in Eq.~(\ref{Gcontract}), i.e.,
\begin{eqnarray}
W_{i_{1}i_{2};i_{3}'i_{4}'}
=
\sum_{i_{5}i_{6}}
X_{i_{1}i_{2};i_{5}i_{6}}
Y_{i_{5}i_{6};i_{3}'i_{4}'}
\times
(-1)^{p(i_{6})}
=
\sum_{i_{5}i_{6}}
X_{i_{1}i_{2};i_{5}i_{6}}
\tilde{Y}_{i_{5}i_{6};i_{3}'i_{4}'}
\label{Wdef}
\end{eqnarray}
where we have absorbed the sign factor $(-1)^{p(i_{6})}$ into $\tilde{Y}$. We choose $\mathcal{P}$ and $\mathcal{Q}$ in such a way that they constitute the truncated Grassmann singular value decomposition (SVD) of $\mathcal{W}$ as shown in Fig. \ref{ctmrg_isometry}:
\begin{eqnarray} \label{goal}
\mathcal{W}_{\Psi_{1}\Psi_{2};\bar{\Psi}_{3}\bar{\Psi}_{4}}
\approx
\int
\mathcal{U}_{\Psi_{1}\Psi_{2};\Phi_{1}}
\mathcal{S}_{\bar{\Phi}_{1}\Phi_{2}}
\mathcal{V}^{\dagger}_{\bar{\Phi}_{2};\bar{\Psi}_{3}\bar{\Psi}_{4}}
=
\int
\mathcal{X}_{\Psi_{1}\Psi_{2};\Psi_{5}^{l}\bar{\Psi}_{6}^{l}}
(
\mathcal{P}_{\Psi_{6}^{l}\bar{\Psi}_{5}^{l};\bar{\Psi}_{7}}
\mathcal{Q}_{\Psi_{7};\Psi_{5}^{r}\bar{\Psi}_{6}^{r}}
)
\mathcal{Y}_{\Psi_{6}^{r}\bar{\Psi}_{5}^{r};\bar{\Psi}_{3}\bar{\Psi}_{4}}
\end{eqnarray}
where we have used $\int$ as a shorthand notation for all the Grassmann integral measures, and the dimensions corresponding to indices $\{\Phi,\bar\Phi\}$ are truncated to $\chi$. It is easy to prove that the solution of $\mathcal{P}$ and $\mathcal{Q}$ in Eq.~(\ref{goal}) can be obtained as
\begin{eqnarray}
& & \label{P}
\mathcal{P}_{\Psi_{6}\bar{\Psi}_{5};\bar{\Psi}_{7}}
=
\sum_{i_{5}'i_{6}i_{7}'}
\left(
P_{i_{5}'i_{6};i_{7}'}
\times
(-1)^{p(i_{6})}
\right)
\Psi_{6}^{p(i_{6})}
\bar{\Psi}_{5}^{p(i_{5}')}
\bar{\Psi}_{7}^{p(i_{7}')}
\\ 
& & \label{Q}
\mathcal{Q}_{\Psi_{7};\Psi_{5}\bar{\Psi}_{6}}
=
\sum_{i_{5}i_{6}'i_{7}}
\left(
Q_{i_{7};i_{5}i_{6}'}
\times
(-1)^{p(i_{7})}
\right)
\Psi_{7}^{p(i_{7})}
\Psi_{5}^{p(i_{5})}
\bar{\Psi}_{6}^{p(i_{6}')}
\end{eqnarray}
where the coefficient tensors $P$ and $Q$ are defined in the following:
\begin{eqnarray} \label{PQ}
P = \tilde{Y}VS^{-1/2},
\quad
Q = S^{-1/2}U^{\dagger}X
\end{eqnarray}

To see this, using the fact $USV^{\dagger} = X\tilde{Y}$, we plug Eqs.~(\ref{P}-\ref{PQ}) into the right-hand side of Eq.~(\ref{goal}):
\begin{eqnarray} \label{proof1}
\mathcal{W}_{\Psi_{1}\Psi_{2};\bar{\Psi}_{3}\bar{\Psi}_{4}}
&\approx&
\int_{\bar{\Psi}_{5}^{l}\Psi_{5}^{l}}
\int_{\bar{\Psi}_{6}^{l}\Psi_{6}^{l}}
\int_{\bar{\Psi}_{7}\Psi_{7}}
\int_{\bar{\Psi}_{5}^{r}\Psi_{5}^{r}}
\int_{\bar{\Psi}_{5}^{r}\Psi_{5}^{r}}
\mathcal{X}_{\Psi_{1}\Psi_{2};\Psi_{5}^{l}\bar{\Psi}_{6}^{l}}
\mathcal{P}_{\Psi_{6}^{l}\bar{\Psi}_{5}^{l};\bar{\Psi}_{7}}
\mathcal{Q}_{\Psi_{7};\Psi_{5}^{r}\bar{\Psi}_{6}^{r}}
\mathcal{Y}_{\Psi_{6}^{r}\bar{\Psi}_{5}^{r};\bar{\Psi}_{3}\bar{\Psi}_{4}}
\\ \nonumber
&=&
\sum_{i_{1}i_{2}i_{3}'i_{4}'}
\left(
\sum_{j_{5}j_{6}i_{7}k_{5}k_{6}}
X_{i_{1}i_{2};j_{5}j_{6}}
P_{j_{5}j_{6};i_{7}}
Q_{i_{7};k_{5}k_{6}}
\tilde{Y}_{k_{5}k_{6};i_{3}'i_{4}'}
\right)
\Psi_{1}^{p(i_{1})}
\Psi_{2}^{p(i_{2})}
\bar{\Psi}_{3}^{p(i_{3}')}
\bar{\Psi}_{4}^{p(i_{4}')}
\end{eqnarray}
where the sign factor from the Grassmann contraction of $\mathcal{X}$ and $\mathcal{P}$ cancels that in $\mathcal{P}$, the sign factor from contracting $\mathcal{P}$ and $\mathcal{Q}$ cancels that in $\mathcal{Q}$, and the sign factor from the Grassmann contraction of $\mathcal{Q}$ and $\mathcal{Y}$ is incorporated in $\tilde{Y}$. Since the coefficients satisfy the following equation
\begin{eqnarray}
XPQ\tilde{Y} 
&=& 
X
\left( \tilde{Y}VS^{-1/2} \right)
\left( S^{-1/2}U^{\dagger}X \right)
\tilde{Y}
=
\left( 
X
\tilde{Y}
\right)
V
S^{-1/2}
S^{-1/2}
U^{\dagger}
\left( X 
\tilde{Y}
\right)
\\ \nonumber
&=&
\left(
U
S
V^{\dagger}
\right)
V
S^{-1}
U^{\dagger}
\left(
U
S
V^{\dagger}
\right)
=
U
S
\left(
V^{\dagger}
V
\right)
S^{-1}
\left(
U^{\dagger}
U
\right)
S
V^{\dagger}
\\ \nonumber
&=&
USV^{\dagger}
\end{eqnarray}
thus the equation below, corresponding to Eq.~(\ref{goal}), holds:
\begin{eqnarray}
\mathcal{W}_{\Psi_{1}\Psi_{2};\bar{\Psi}_{3}\bar{\Psi}_{4}}
\approx
\sum_{i_{1}i_{2}i_{3}'i_{4}'}
\left(
USV^{\dagger}
\right)_{i_{1}i_{2};i_{3}'i_{4}'}
\Psi_{1}^{p(i_{1})}
\Psi_{2}^{p(i_{2})}
\bar{\Psi}_{3}^{p(i_{3}')}
\bar{\Psi}_{4}^{p(i_{4}')}
=
\int
\mathcal{U}_{\Psi_{1}\Psi_{2};\Phi_{1}}
\mathcal{S}_{\bar{\Phi}_{1}\Phi_{2}}
\mathcal{V}^{\dagger}_{\bar{\Phi}_{2};\bar{\Psi}_{3}\bar{\Psi}_{4}}
\end{eqnarray}

The sign factors appeared in Eqs.~(\ref{P}-\ref{Q}) might be slightly different in constructing the Grassmann isometries for moves in the other three directions, but the methodology is the same. Once the Grassmann CTMRG procedure is converged, with the help of converged environment tensors \cite{VCTMRG2022}, the desired physical quantities can be easily obtained through the impurity method mentioned in Sec.\ref{Sec:GTNS}. In this context, the expectation value of any local observable can be expressed as a ratio of two tensor network scalars, as expressed in Eq.~(\ref{ctmrg_exp}).
\begin{eqnarray} \label{ctmrg_exp}
\langle 
\hat{O}^{\textrm{imp}}
\rangle
=
\raisebox{-0.48\height}{\includegraphics[width=0.3\textwidth, page=1]{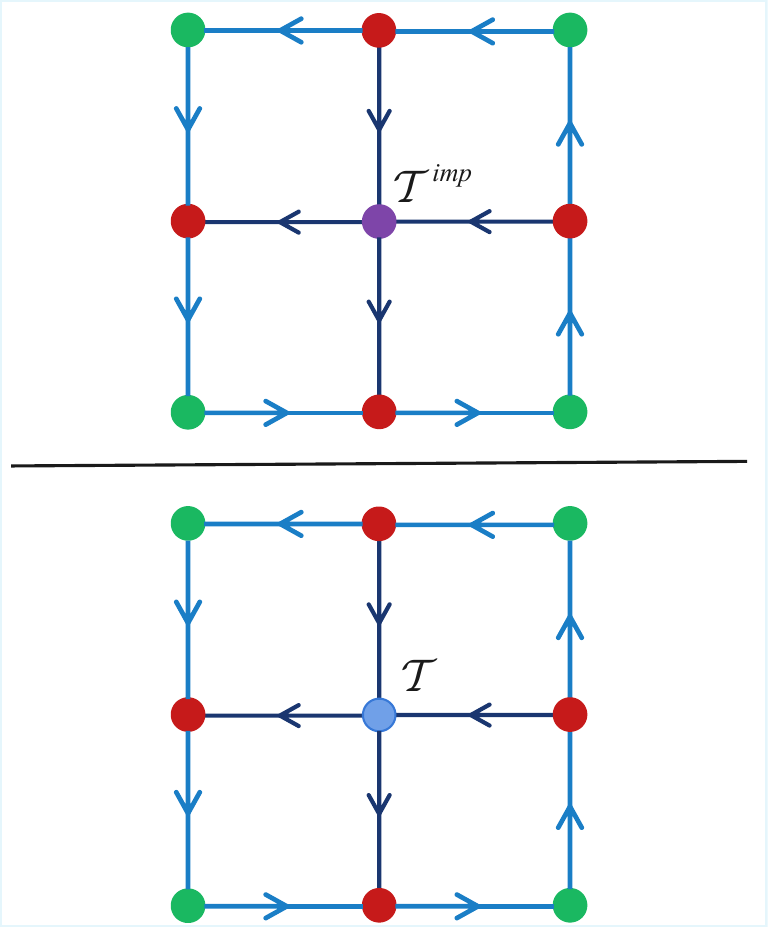}}
\end{eqnarray} 
where the environment tensors surrounding $\mathcal{T}$ and $\mathcal{T}^{\textrm{imp}}$ are obtained from the Grassmann CTMRG procedure described above.

\subsection*{Details for numerical implementation}

In this work, we choose a tiny discretization step in the imaginary-time direction, $\epsilon = 10^{-4}$, to reduce the Trotter error at the initial stage as much as possible. A point is that in the Grassmann CTMRG method, the inverse temperature $\beta$ grows only linearly with the iteration step, thus if we rely solely on this method to address the ground state properties of a given model, with such a small $\epsilon$, the required renormalization steps to approach the zero-temperature limit would be huge. Therefore, we prefer to utilize the Grassmann high-order tensor renormalization group (HOTRG) algorithm \cite{Xie2012,Yuya2014,Akiyama2021,Akiyama2021a} as a pre-processing tool to provide better initial Grassmann tensors for the Grassmann CTMRG method. 

In a recent work \cite{Akiyama2021}, a hyperparameter $N_{\tau}$ is introduced to control the Grassmann HOTRG steps in the temporal direction before performing the ordinary iterations. It serves as an effective strategy to reduce the anisotropy in the Grassmann tensor caused by the small $\epsilon$, and a similar idea has been first used to study the ground state of the 2D quantum Ising model \cite{Xie2012}. We adopt the same strategy with a fixed bond dimension $D_{c}=16$ for the HOTRG procedure and find that $N_{\tau} = 8$ yields negligible errors in the Grassmann HOTRG procedure and works well in most cases. In addition to reduce the anisotropy in tensor elements, the effective temperature corresponding to the Grassmann tensor after the $N_{\tau}$ temporal steps would become $\beta_{0} = 2^{N_{\tau}}\times\epsilon = 0.0256$, thus largely reducing the required Grassmann CTMRG iterations to approach zero temperature. For example, after 8000 Grassmann CTMRG iterations, we find that the convergence error of the expectation values is already lower than $10^{-8}$ for $\chi=64$. 

Furthermore, we find that the following blocking operation (without truncation) along the spatial direction, after the $N_{\tau}$ HOTRG steps in the imaginary-time direction, could also enhance the numerical stability for the Grassmann CTMRG:
\begin{eqnarray} \label{grouping}
\raisebox{-0.49\height}{\includegraphics[width=0.9\textwidth, page=1]{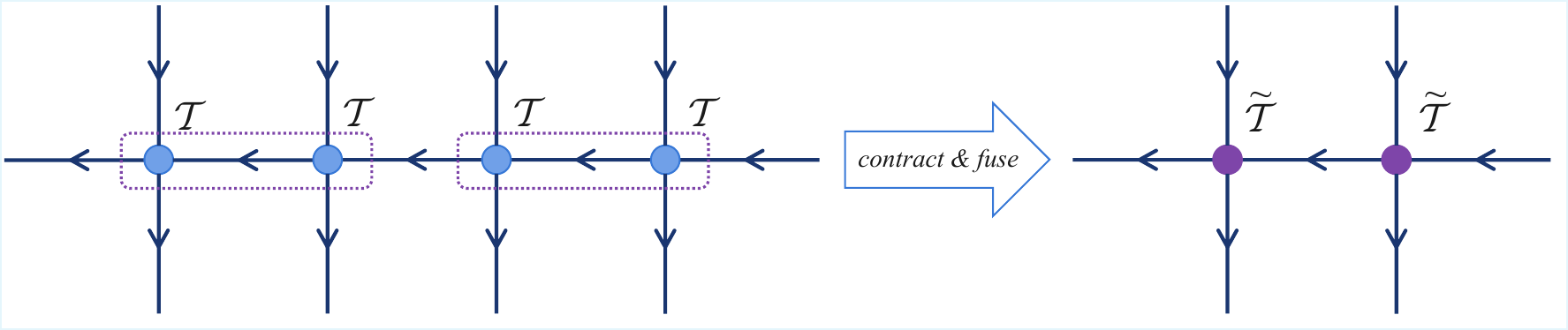}}
\end{eqnarray}
That is, we perform a single step of Grassmann contraction between the neighbouring Grassmann tensors along the spatial direction, and fuse the two temporal indices into a single one.

Hereafter, for clarity, we call the modified method with the pre-processing procedure by Grassmann HOTRG the GCTMRG* method. For the one-dimensional Hubbard model expressed in Eq.~(\ref{hubbard_H}), the GCTMRG* is performed in the following steps: (i) given the initial Grassmann tensor $\mathcal{T}_{0}$ defined in Eqs.(\ref{gtensor_bulk}-\ref{coef_tensor_bulk}), which is of size $16\times4\times16\times4$, we perform $N_{\tau}=8$ Grassmann HOTRG steps only in the imaginary-time direction with a cutoff $D_{c} = 16$, and obtain the renomalized tensor $\mathcal{T}_{1}$ of the same size as the initial $\mathcal{T}_0$; (ii) we perform a single step of the blocking operation illustrated in Eq.~(\ref{grouping}) and obtain a Grassmann tensor $\mathcal{T}_{2}$ of size $16\times16\times16\times16$; (iii) the Grassmann CTMRG algorithm, discussed in this section, is performed to approximate the environments of $\mathcal{T}_{2}$ in the zero-temperature limit; (iv) the expectation values are evaluated with the help of the converged environments from (iii), as illustrated in Eq.~(\ref{ctmrg_exp}).

\section{Results}
\label{Sec:Res}
\begin{figure}[htbp]
    \centering
    \includegraphics[height=8.8cm,width=13.0cm]{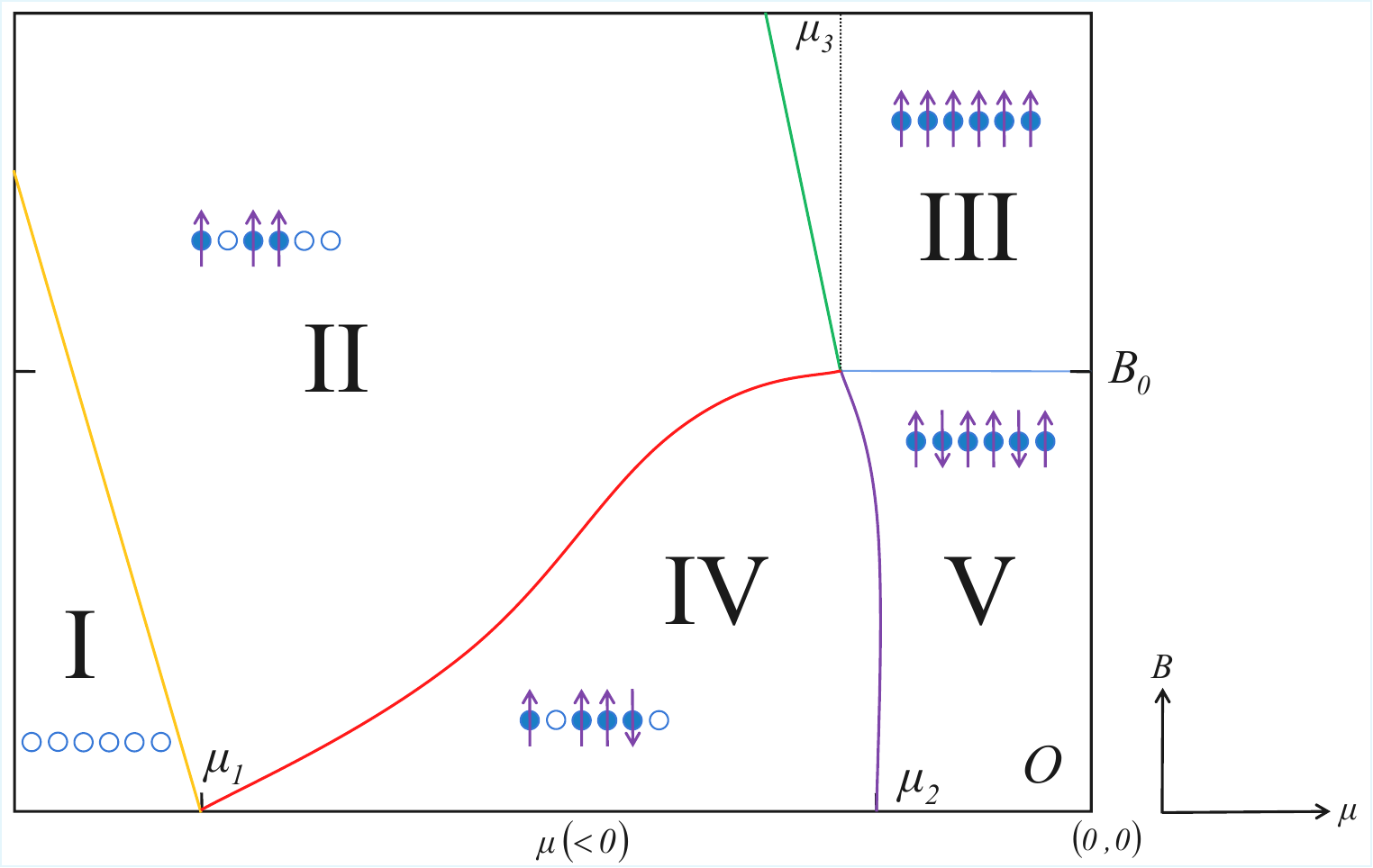}
    \caption{The phase diagram of the one-dimensional Hubbard model in the $(\mu, B)$ plane, obtained from the Bethe Ansatz. This figure is adapted from Ref.~\cite{Essler2005}. Schematic plots of each phase are provided to highlight their key characteristics. In these illustrations, an empty circle denotes an unoccupied lattice site, and a filled circle with an arrow represents an electron with the arrow indicating the spin orientation. $\mu_{1}$ and $\mu_{2}$ are defined in the main text, $\mu_{3} = 2 - 2\sqrt{1+u^{2}}$, $B_{0} = 2\sqrt{1 + u^2} - 2u$ with $u \equiv U/4$, and $U$ denotes the on-site interaction strength. For $U=5$, a value used in part of our numerical computations, $B_{0} \approx 0.702$, and $\mu_{3} \approx -1.202$.}
    \label{phase_diagram_full}
\end{figure}

The ground state phase diagram of the one-dimensional Hubbard model in the $(\mu,B)$ plane can be obtained using the Bethe Ansatz method \cite{Essler2005}. A schematic plot of the phase diagram with $\mu < 0$ and $B > 0$ is shown in Fig.~\ref{phase_diagram_full}. In particular, when $B=0$, the ground state of the model can be analytically solved, thus providing a perfect platform to test our GCTMRG* method. 

As shown in Fig.~\ref{phase_diagram_full}, the phase diagram is divided into five distinct parts. The phase I stands for a completely trivial phase corresponding to an empty lattice with $p = 0$, in the extremely negative $\mu$ region. The phase II corresponds to partially-filled states with electron densities $0 < p < 1$, while the spins are fully polarized, i.e., net magnetization density $m = p/2$. Phase III, located in the upper right part of the phase diagram, corresponds to the half-filled, fully-polarized ground state, namely $p=1$ and $m=1/2$. The phase IV is an intermediate metallic phase characterized by partially-filled electrons as well as partially-polarized spins, i.e., $0<p<1$ and $0<m<p/2$. Finally, the phase V denotes an insulating ground state with half-filled and magnetic bands, that is, $p=1$ and $0<m<p/2$.

\begin{figure}[htbp]
	\centering
	\begin{minipage}{0.48\textwidth}
		\centering
		\includegraphics[height=5.0cm,width=7.6cm]{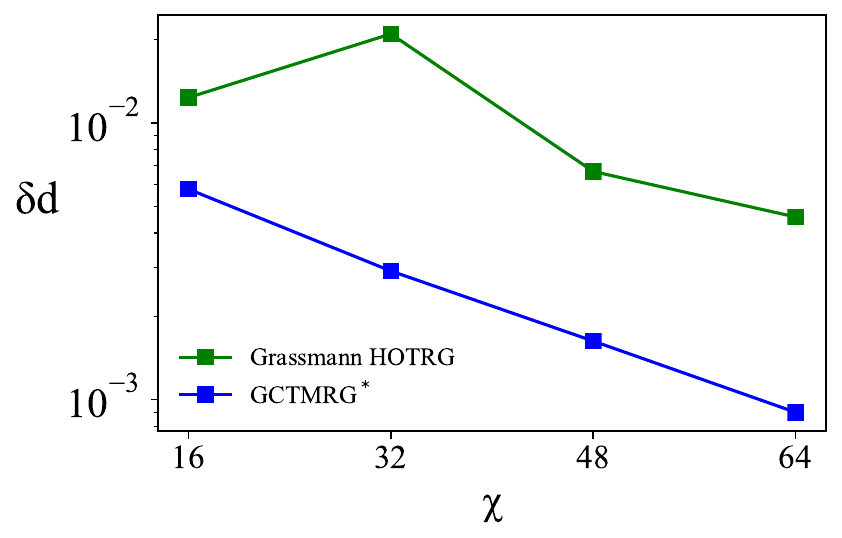}
	\end{minipage}
	\begin{minipage}{0.48\textwidth}
		\centering
		\includegraphics[height=5.0cm,width=7.6cm]{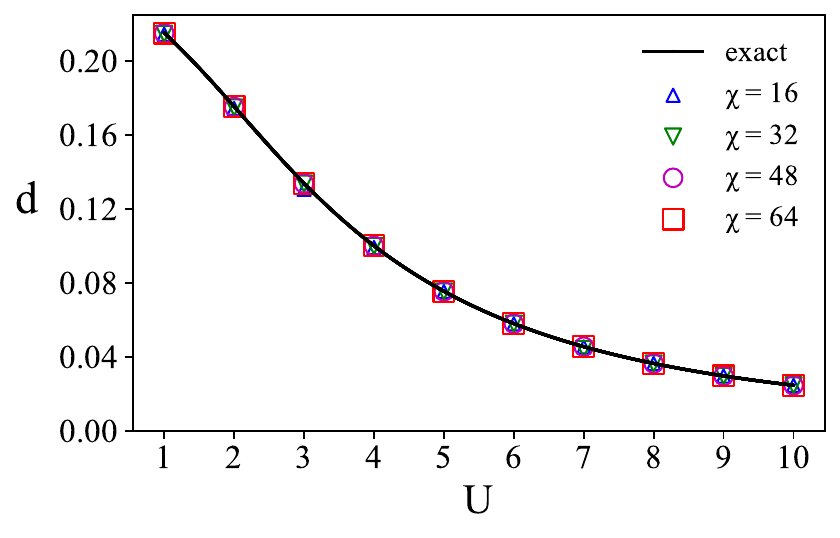}
	\end{minipage}
	\caption{Performance of the double occupancy $d=\langle\hat{n}_{\uparrow}\hat{n}_{\downarrow}\rangle$ for the one-dimensional Hubbard model at $(0, 0)$ point in the phase diagram. The exact results expressed in Eq.~(\ref{exact}) are also included for comparison. (a) Relative error $\delta d$ with respect to the exact solution at $U = 2$. Performance of Grassmann HOTRG and GCTMRG* obtained at a series of bond dimension $\chi$ is compared. (b) $d$ as a function of $U$. The data are obtained from the GCTMRG* method with a series of $\chi$.}
	\label{mu0B0}
\end{figure}

We firstly focus on the origin point of the $(\mu, B)$ plane, i.e., $O(0, 0)$ located at lower right corner of the phase diagram in Fig.~\ref{phase_diagram_full}, validate our GCTMRG* method, and highlight the its improvements over the Grassmann HOTRG approach by computing the relative error of the double occupancy $d = \langle\hat{n}_\uparrow\hat{n}_\downarrow\rangle$ in the ground state with respect to the exact solution,
\begin{eqnarray} \label{exact}
d = \dfrac{1}{2} \int_{0}^{\infty} d\omega
\mathrm{J}_{0}(\omega)
\mathrm{J}_{1}(\omega)
\operatorname{sech}^2(\omega U/4) 
\end{eqnarray}
where $U$ denotes the on-site Hubbard interaction strength, and the $\mathrm{J}_{0}(\omega)$ and $\mathrm{J}_{1}(\omega)$ are the zeroth-order and first-order Bessel functions of the first kind, respectively. For Grassmann HOTRG, the double occupancy $d$ can be obtained via 
\begin{eqnarray} \label{dc}
d
-
\dfrac{p}{2}
+
\dfrac{1}{4}
=
- 
\dfrac{1}{\beta L}
\dfrac{\partial \ln(Z)}{\partial U} \bigg|_{\mu,B,T}
\end{eqnarray}
where $L$ is the length of the chain. Here, $Z$ denotes the partition function defined in Eq.~\eqref{hubbard_partition_function}, which is a function of parameters $U$, $\mu$, $B$, and $T$. The relation above can be verified directly by substituting $Z = \textrm{Tr}\left(e^{-\beta \hat{H}}\right)$ and the explicit Hamiltonian in Eq.~\eqref{hubbard_H} into the right-hand side of Eq.~\eqref{dc}.

For the $(0, 0)$ point, the particle density $p = 1$ in Eq.~(\ref{dc}) due to the particle-hole symmetry. In Grassmann HOTRG, the partition function $Z$ can be obtained directly, and its finite difference yields $d$ according to Eq.~(\ref{dc}). In this work, setting the Trotter step $\epsilon=10^{-4}$, we have chosen pre-processing steps $N_{\tau}=12$, and then $N=24$ steps of Grassmann HOTRG are performed to approach the zero-temperature limit \cite{Akiyama2021}. The finite difference is performed with $\Delta U = 10^{-4}$. While in the GCTMRG* method, once the environment tensors are obtained, $d$ can be estimated conveniently by the impurity method according to Eqs.~(\ref{nund_impurity}) and (\ref{ctmrg_exp}). To avoid confusion, we adopt the following convention: in the pre-processing stage of the GCTMRG* algorithm, the Grassmann HOTRG bond dimension is denoted by $D_{c}$, whereas for standard Grassmann HOTRG, it is denoted by $\chi$, the same as the Grassmann CTMRG. The $D_{c}$ is fixed at 16 throughout our numerical computations.

The result is summarized in Fig.~\ref{mu0B0}. In Fig.~\ref{mu0B0}(a), the double occupancy $d$ obtained for $U = 2$ are compared. We can see the the GCTMRG* method outperforms the Grassmann HOTRG at a series of environment dimensions $\chi$. Besides, as shown in Fig.~\ref{mu0B0}(b), the double occupancy obtained from the GCTMRG* decreases with increasing $U$ as expected, and the numerical value is in excellent agreement with the exact solutions in Eq.~(\ref{exact}) across a wide range of on-site interactions $U$. Moreover, \ the data obtained from different $\chi$ exhibit perfect convergence, and this verifies that our GCTMRG$^*$ method can efficiently capture ground state physics in the one-dimensional Hubbard model in the thermodynamic limit.

Once the validity of our approach is established, we now turn to explore the whole phase diagram by calculating the order parameters, i.e., the density of particle number $p = \langle \hat{n}_{\uparrow} \rangle + \langle \hat{n}_{\downarrow} \rangle$ and the magnetization $m = \left(\langle \hat{n}_{\uparrow} \rangle - \langle \hat{n}_{\downarrow} \rangle\right)/2$, with the Grassmann tensor network approach along three typical lines in the $(\mu, B)$ plane. In the GCTMRG* calculations, we set $\chi = 64$ for the Grassmann CTMRG iterations hereafter. 

The first line we focus on is $B = 0$ in Fig.~\ref{phase_diagram_full}. In this case, the corresponding one-dimensional phase diagram as a function of chemical potential $\mu$ is
\begin{equation} \label{phase_diagram_B0}
\raisebox{-0.4\height}{\includegraphics[width=0.44\textwidth, page=1]{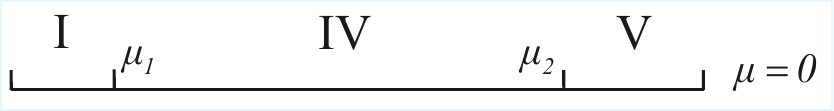}}
\end{equation}
where both the critical value $\mu_{1}$ separating the trivial phase ($p=0$) and the intermediate metalliclic phase $(0<p<1)$, as well as the critical $\mu_{2}$ separating the metallic phase and the insulating phase $(p=1)$ can be analytical determined by \cite{Essler2005}
\begin{equation} \label{mu_critical}
\mu_{1}(u)=-2-2u,
\qquad
\mu_{2}(u) = 2 - 2u - 2\int_{0}^{\infty} 
\dfrac{d\omega}{\omega}\dfrac{\mathrm{J}_{1}(\omega)\mathrm{e}^{-\omega u}}{\cosh{(\omega u)}}, 
\quad
u \equiv U/4
\end{equation}
\begin{figure}[htbp] 
    \centering
    \includegraphics[height=5.8cm,width=8.4cm]{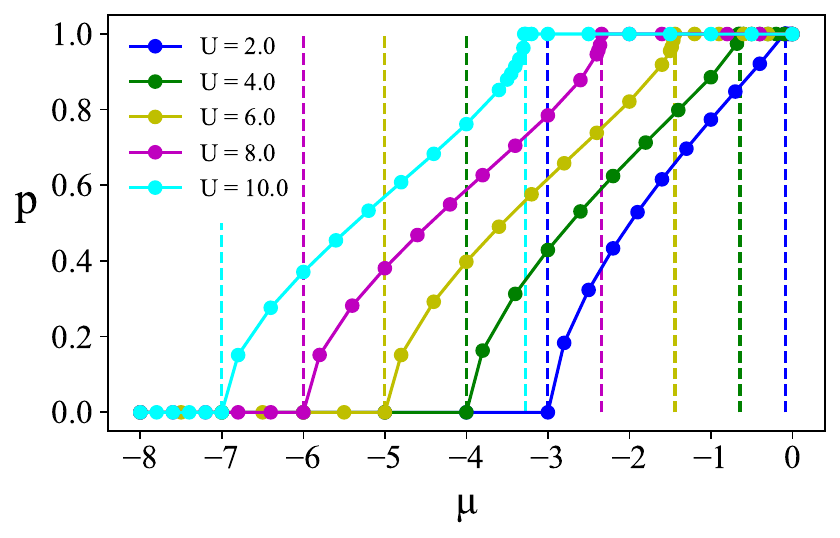}
    \caption{Particle number $p = \langle \hat{n}_{\uparrow} \rangle + \langle \hat{n}_{\downarrow} \rangle$ obtained by GCTMRG*, as a function of chemical potential $\mu$, for the one-dimensional Hubbard model with a series of $U$ and $B = 0$. The exact values of the critical $\mu$ at the phase transitions for these $U$, determined by Eq.~(\ref{mu_critical}), are also included as dashed lines for comparison.}
    \label{n_B0}
\end{figure} 

The particle number $p$ obtained from the GCTMRG* method is shown in Fig.~\ref{n_B0}. It shows that for a given $U$ and an increasing $\mu$, as expected, the particle density starts from zero in the trivial empty phase I, gradually becomes larger in the intermediate metallic phase IV, and finally reaches one in the insulating phase V. In fact, the obtained critical chemical potentials indicating the phase transitions are also consistent with the exact values expressed in Eq.~(\ref{mu_critical}): for $U=2, 4, 6, 8, 10$, we have $\mu_{1} = -3, -4, -5, -6, -7$, and $\mu_{2} \approx -0.086, -0.643, -1.446, -2.339, -3.273$, which are denoted as dashed lines in Fig.~\ref{n_B0} for comparison.

\begin{figure}[htbp]
	\centering 
	\begin{minipage}{0.325\textwidth}
		\centering 
		\begin{overpic}[height=4.0cm,width=5.6cm]{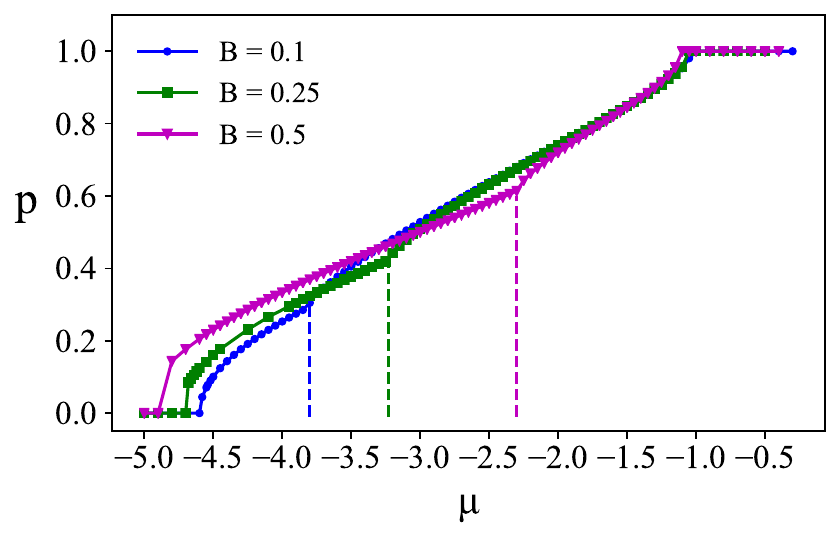}
			\put(-2,71){(a)}
		\end{overpic}
	\end{minipage}
	\begin{minipage}{0.325\textwidth}
		\centering 
		\begin{overpic}[height=4.0cm,width=5.6cm]{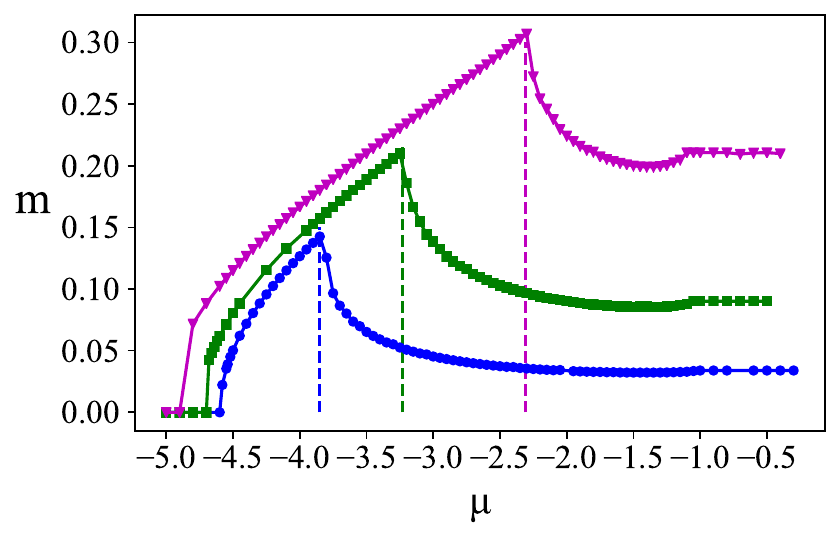}
			\put(-2,71){(b)}
		\end{overpic}
	\end{minipage}
	\begin{minipage}{0.325\textwidth}
		\centering 
		\begin{overpic}[height=4.0cm,width=5.6cm]{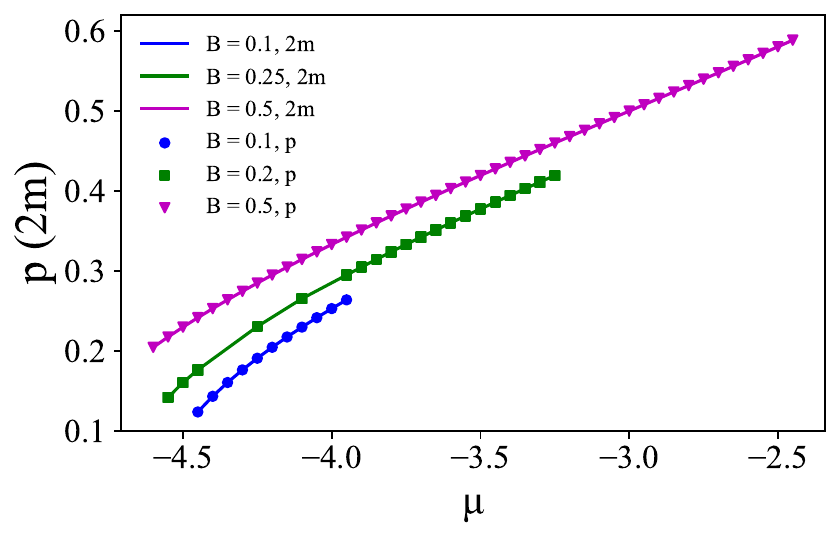}
			\put(-2,71){(c)}
		\end{overpic}
	\end{minipage}
	\caption{(a) Particle number $p$ and (b) magnetization $m = \left(\langle \hat{n}_{\uparrow} \rangle - \langle \hat{n}_{\downarrow} \rangle\right)/2$, obtained by the GCTMRG* method, as functions of chemical potential $\mu$, for a series of magnetic fields $B$ and a fixed $U = 5$. The dashed lines indicating the phase transitions between phases II and IV are added for guidance. (c) The relation $p = 2m$ within phase II is verified.}
	\label{U5ab}
\end{figure}

Next, we turn on the external fields $B = 0.1, 0.25, 0.5$ with $U$ fixed at 5. In this case, the considered magnetic fields are below $B_{0} \approx 0.702$, the corresponding phase diagram should look like
\begin{equation}
\raisebox{-0.4\height}{\includegraphics[width=0.44\textwidth, page=1]{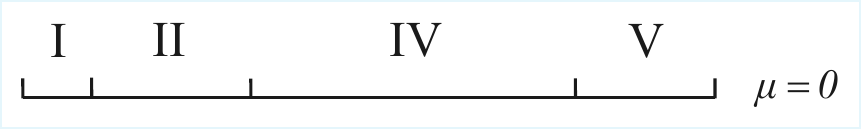}}
\end{equation}
Compared to Eq.~(\ref{phase_diagram_B0}), an additional phase, i.e., phase II with fully-polarized spin and partially-filled state $(0<p<1, m=p/2)$, emerges between the trival phase I and the  phase IV. The obtained results of $p$ and $m$ from GCTMRG* are shown in Fig.~\ref{U5ab}. Here, we focus on the line $B = 0.25$ for illustration. For an extremely negative $\mu$, the trivial phase with zero density and magnetization is verified. As $\mu$ increases, we enter the fully-polarized phase II, where the expected relation $m = p/2$ is numerically confirmed, see Fig.~\ref{U5ab}(c). The transition between the phase II and the phase IV can be inferred from either the cusp in the plot of $p(\mu)$ shown in Fig. \ref{U5ab}(a), or the peak of the $m(\mu)$ shown in Figure \ref{U5ab}(b). Finally, the phase V with $p=1$ is obtained when $\mu$ further increases. Similar conclusions can also be drawn from calculations with $B = 0.1$ and $B = 0.5$, as shown in Fig.~\ref{U5ab}.

\begin{figure}[htbp]
	\centering 
	\begin{minipage}{0.48\textwidth}
		\centering 
		\begin{overpic}[height=5.2cm,width=7.6cm]{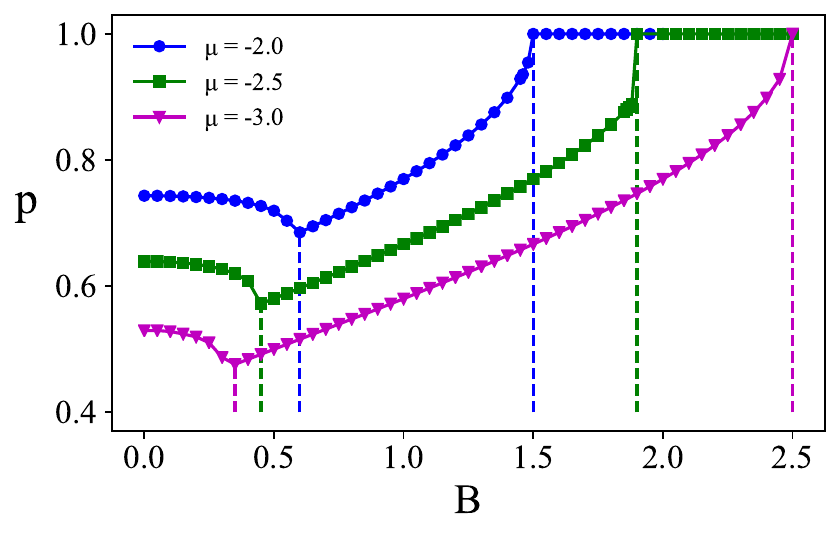}
			\put(-2,60){(a)}
		\end{overpic}
	\end{minipage}
	\begin{minipage}{0.48\textwidth}
		\centering 
		\begin{overpic}[height=5.2cm,width=7.6cm]{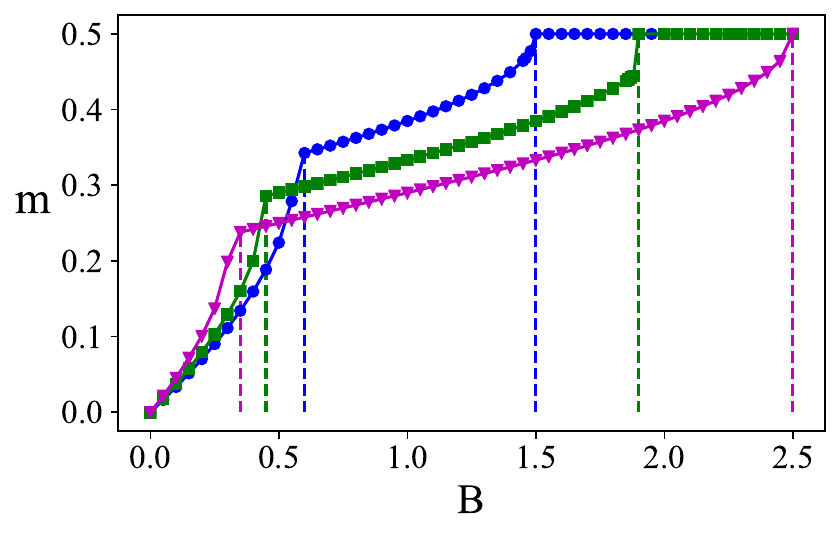}
			\put(-2,60){(b)}
		\end{overpic}
	\end{minipage}
	\caption{(a) Particle number and (b) magnetization as functions of the magnetic field $B$, for a series of chemical potentials. $U = 5$ is fixed in the calculations. The dashed lines indicating the phase transitions are added for guidance.}
	\label{mu-234ab}
\end{figure}

Finally, we perform calculations along the lines $\mu = -2.0, -2.5, -3.0$, which is smaller than $\mu_{2}(u=1.25)\approx -1.03$ for $U = 5$, and the corresponding phase diagram should look like
\begin{equation} \label{phase_diagram_varyB}
\raisebox{-0.4\height}{\includegraphics[width=0.44\textwidth, page=1]{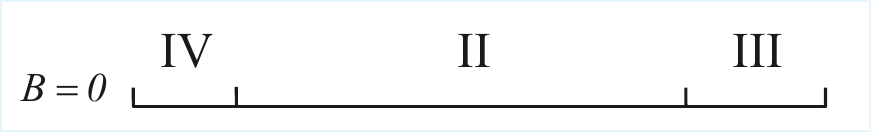}}
\end{equation}
As illustrated in Fig.~\ref{mu-234ab}, if we focus on $\mu = -2.5$ for the moment, the system starts from the intermediate phase IV with zero magnetization, where as $B$ grows, the particle number decreases gradually, and the magnetization becomes larger. Then the system transitions to the fully-polarized phase II, as also evidenced by the relation $m \approx p/2$. With an even larger magnetic field, the model finally enters the half-filled, fully-polarized phase III, characterized by $p = 1$ and $m = 1/2$. All these features are consistent with the phase diagram illustrated in Fig.~\ref{phase_diagram_full} and Eq.~(\ref{phase_diagram_varyB}). For other chemical potentials like $\mu = -2$ and $\mu = -3$, we have the same conclusion.

With all these results, we conclude that the GCTMRG* method based on the Grassmann tensor networks could indeed reproduce the key features in the phase diagram of the one-dimensional Hubbard model under the magnetic field, demonstrating its ability to study the one-dimensional interacting fermionic models.

\section{Summary}
\label{Sec:Summary}

This work presents a Grassmann CTMRG method assisted by the Grassmann HOTRG steps \cite{Xie2012,Yuya2014,Akiyama2021,Akiyama2021a}, dubbed the GCTMRG* method, to study the ground-state properties of one-dimensional interacting fermionic models. Representing the partition function of a lattice fermion in the path-integral formulation as a (1+1)-dimensional Grassmann tensor network, the method treats the fermionic model just like a 2D anisotropic classical model, but with Grassmann variables involved. The Grassmann HOTRG serves as a pre-processing tool to reduce the anisotropy of the Grassmann tensor network and efficiently reduces the required renormalization steps in the imaginary-time direction in the subsequent Grassmann CTMRG iterations. The GCTMRG* method demonstrates its advantages over the direct Grassmann HOTRG method through the double-occupancy calculations for the one-dimensional Hubbard model with $\mu=0$ and $B=0$. More importantly, we show that our method captures the essential features of the full phase diagrams, demonstrating its ability to investigate the one-dimensional interacting fermionic model. 

In this work, the main goal of the calculations is to test the validity of the Grassmann CTMRG algorithm in characterizing the full phase diagram, as shown in Fig.~\ref{phase_diagram_full}. With $D_c = 16$, $\chi = 64$, we have shown that the phase transitions detected clearly in Figs.~\ref{mu0B0}, \ref{n_B0}, \ref{U5ab}, and \ref{mu-234ab} are sufficient to reconstruct the phase diagram. Furthermore, one can use a larger $D_c$, a larger $\chi$, and smaller parameter intervals (such as $\Delta\mu$ and $\Delta B$) to refine the phase diagrams by determining the phase boundaries more accurately. 

As a prospect, it would be interesting to explore the possibilities for finite-temperature computations \cite{Li2011,Dong2017} of one-dimensional fermionic models in the coherent-state path-integral formalism, where a Grassmann version of the time-evolving block decimation method \cite{Orus2008} is necessary. It is also worthwhile to compare the two routes to interacting fermion models in some details, i.e., the current Grassmann tensor network approach based on the path-integral representation of the partition functions \cite{Yuya2014, Takeda2015, Sakai2017, Yoshimura2018, Akiyama2021a, Akiyama2024}, and the fermionic tensor network states based on the variational principles applied to the variational wave functions satisfying the Fermi-Dirac statistics \cite{Gu2010, Gu2013, Corboz2009, Corboz2010, NickPRB2017, Mortier2025}. At last, while the partition function of a 2D fermionic model represented as a (2+1)-dimensional Grassmann tensor network can be complicated to contract, it remains valuable for the study of the thermodynamics of 2D systems in the thermodynamic limit, to which we hope to contribute in the future.

\addcontentsline{toc}{chapter}{Acknowledgment}
\section*{Acknowledgment}
We are grateful to Dr. Shinichiro Akiyama for helpful discussions. This work was supported by the National Natural Science Foundation of China (Grant No. 12274458) and the National Key R\&D Program of China ( Grant Nos. 2024YFA1408604 and 2023YFA1406500).

\newpage

\end{document}